\documentclass[aps,prd,superscriptaddress,showkeys,showpacs,twocolumn,longbibliography,nofootinbib,preprintnumbers,floatfix]{revtex4-1}
\usepackage[utf8]{inputenc}
\usepackage{amsmath}
\usepackage{amsfonts}
\usepackage{amsthm}
\usepackage{amssymb}
\usepackage{float}
\usepackage{braket}
\usepackage{graphicx}
\usepackage{xcolor}
\usepackage{url}
\usepackage[colorlinks=true, pdfstartview=FitV, linkcolor=red, citecolor=blue, urlcolor=blue]{hyperref}
\usepackage{slashed}
\usepackage{mathtools}
\usepackage{dsfont}

\usepackage[normalem]{ulem}
\usepackage{fancyhdr}
\usepackage[capitalise]{cleveref}
\usepackage{placeins}
\usepackage{subfigure}
\begin{document}

\preprint{IQuS@UW-21-123, NT@UW-26-9}

\title{Entanglement in the $\theta$-vacuum}
\author{Sebastian Grieninger} 
\email{segrie@uw.edu}
\affiliation{InQubator for Quantum Simulation (IQuS), Department of Physics, University of Washington, Seattle, WA 98195}

\author{Dmitri E. Kharzeev} 
\email{dmitri.kharzeev@stonybrook.edu}
\affiliation{
 Center for Nuclear Theory, Department of Physics and Astronomy,
Stony Brook University, Stony Brook, New York 11794-3800, USA
}
\affiliation{Energy and Photon Sciences Directorate, Condensed Matter and Materials Sciences Division,
Brookhaven National Laboratory, Upton, New York 11973-5000, USA}

\author{Eliana Marroquin} 
\email{eliana.marroquin@stonybrook.edu}
\affiliation{
 Center for Nuclear Theory, Department of Physics and Astronomy,
Stony Brook University, Stony Brook, New York 11794-3800, USA
}

\begin{abstract}
We compute the entanglement entropy and the entanglement spectrum of the vacuum state in the massive Schwinger model at a finite $\theta$ angle. By computing two-point correlation functions we show that the correlation length diverges in the critical region.
We clarify the physical origin of entanglement entropy enhancement at $\theta=\pi$ by relating it to the competition between distinct electric-flux vacuum branches. We show that the peak near $\theta=\pi$ persists across the range of masses studied and corresponds to the point of maximal competition between distinct vacuum branches with opposite electric-field orientation, where quantum fluctuations due to fermion pair creation are maximized. While this entropy enhancement is generic, {a pronounced compression of the low-lying entanglement spectrum at $\theta=\pi$ distinguishes the near-critical case ($m/g\simeq0.33$)}. 
Using the Bisognano--Wichmann (BW) theorem, we construct a lattice BW entanglement Hamiltonian and compare it with the exact modular Hamiltonian obtained from the reduced density matrix. We observe agreement in the infrared sector, indicating that the entanglement Hamiltonian is well approximated by a spatially weighted Hamiltonian.
Finally, we discuss possible applications to entanglement in topological insulators and quantum wires. {In this work, we used the chirally rotated formulation of the Schwinger model Hamiltonian. We identify an additive vacuum-energy artifact induced by the discretization of the chirally rotated lattice and derive its exact free-lattice form to subtract it in the continuum limit.}
\end{abstract}

\maketitle

\section{Introduction}
The vacuum structure of gauge theories is strongly influenced by their topological sectors. In four-dimensional Yang-Mills theory, gauge configurations are classified by an integer-valued topological charge associated with the Chern-Simons number, giving rise to a family of distinct classical vacua connected by large gauge transformations \cite{Belavin:1975fg,PhysRevLett.37.172,PhysRevD.17.2717}.
The inclusion of a $\theta$ term introduces a CP-violating phase that weights these sectors in the path integral and modifies the vacuum energy density \cite{PhysRevLett.37.172, PhysRevD.17.2717}.
As emphasized in large-$N_c$ studies of $\theta$-dependent gauge theories, the vacuum energy density may exhibit a multi-branched structure in which distinct vacuum branches compete as a function of $\theta$, and the physical vacuum energy density is obtained from their lower envelope \cite{Witten:1980sp,Witten_1998}. {The realization of this structure depends on the dynamics of the theory, including the quark-mass regime.}

An analogous vacuum structure arises in the Schwinger model \cite{Schwinger:1962tp}, the $(1+1)$-dimensional quantum electrodynamics, where the $\theta$-term is equivalent to a background electric field and the vacuum energy becomes multi-branched \cite{Coleman1976}. This makes the Schwinger model a controlled setting to investigate the interplay between topology and quantum entanglement \cite{Zache_2019,Ikeda2020agk,Chakraborty:2020uhf,Grieninger_2024,Thompson:2021eze,Ikeda2023zil,Florio:2024aix,Florio:2025hoc,Lee:2023urk,Florio:2025xup}.

\begin{figure}[h]
    \centering
    \includegraphics[width=0.8\linewidth]{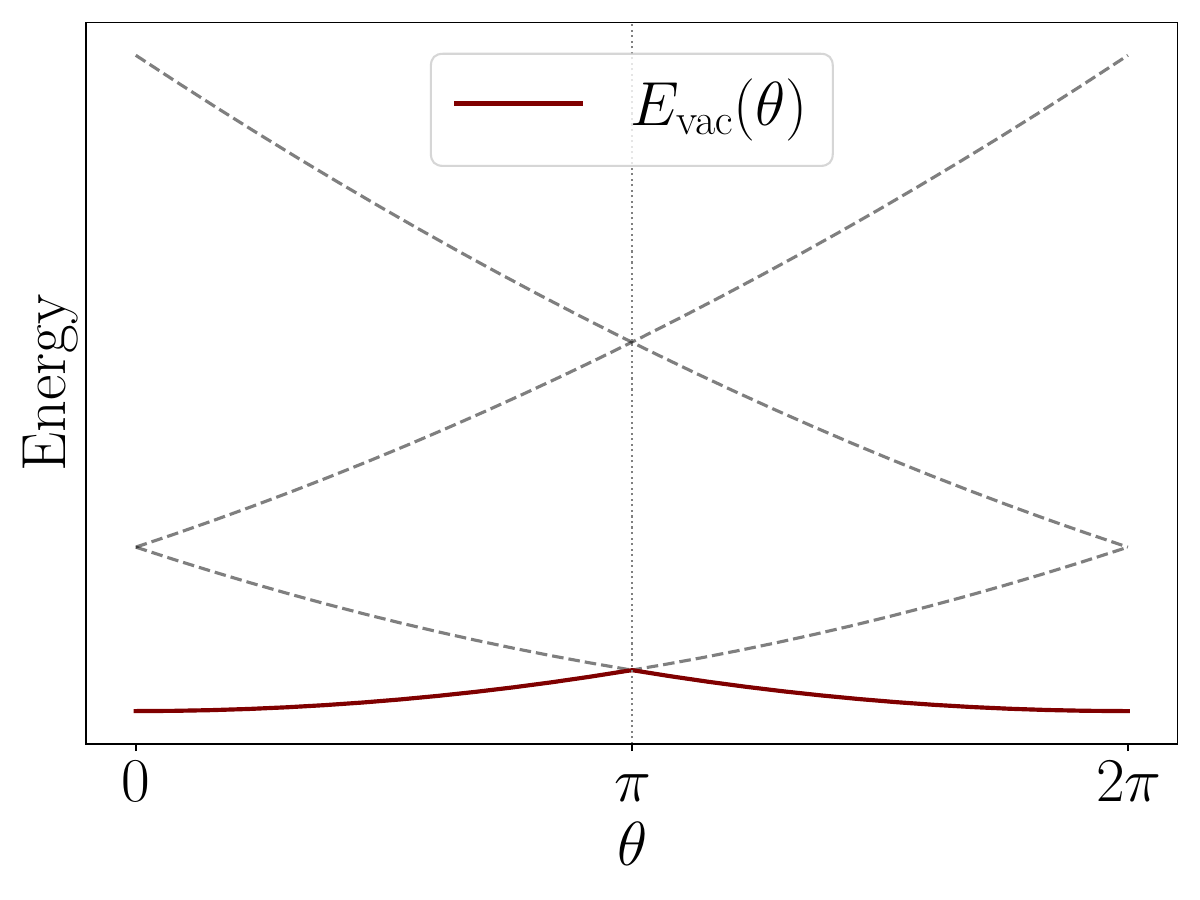}
    \caption{
Schematic vacuum energy density branches $\ E_n(\theta) \!\propto\! (\theta + 2\pi n)^2$ of the Schwinger model in the pure-gauge limit, shown as dashed parabolas. 
The physical vacuum energy $E_{\mathrm{vac}}(\theta)$ (solid red) is obtained by taking the lower envelope of these branches. 
As $\theta$ varies, the system switches between neighboring branches (dashed gray), leading to degeneracy at $\theta = \pi$.
}
    \label{branches}
\end{figure}

In the pure-gauge limit, obtained when the fermions decouple, the vacuum energy consists of electric-flux branches labeled by an integer electric flux number $n$, $ E_n(\theta)\sim (\theta+2\pi n)^2$ . Minimizing over these branches restores $2\pi$ periodicity and produces a branch crossing at $\theta=\pi$, as illustrated in Fig.~\ref{branches}. As the vacuum angle passes $\theta=\pi$, the system switches from one parabola to another. At $\theta=\pi$, two branches with opposite orientations of the background electric field become degenerate, signaling the Dashen phenomenon \cite{PhysRevD.3.1879}. Dynamical fermions modify this structure, in particular, the vacuum energy becomes independent of $\theta$ in the massless continuum theory.

In the massive Schwinger model, bosonization yields an effective potential consisting of a quadratic confining term and a periodic cosine term whose relative strength depends on the fermion mass and gauge coupling. The interplay of these contributions generates not only the true ground state but also a family of metastable branches that appear as displaced local minima of the potential. For general $\theta \neq 0,\pi$, the cosine term induces asymmetry in the effective potential, shifting the location of its local minima away from the parity-even ground state.

Related phenomena appear in effective descriptions of QCD, such as nonlinear sigma models with a $\theta$-term \cite{Kharzeev_1998}, where local minima arise from the interplay between spontaneous chiral symmetry breaking and explicit topological CP-violating terms. While the microscopic origin and the physical interpretation of these minima can differ across models, these examples show that $\theta$-dependent gauge theories have a rich vacuum structure beyond the global ground state. Motivated by this broader context, we investigate how entanglement observables diagnose the $\theta$-dependent vacuum structure across the $\theta$ range and varying fermion mass $m$ and coupling $g$.

\emph{Quantum entanglement} has emerged as a powerful diagnostic of vacuum structure, phase transitions, and topological order. Entanglement measures such as the von Neumann entropy (EE) and entanglement spectrum (ES) were analytically studied in conformal field theory \cite{Holzhey_1994, Pasquale_Calabrese_2004} and play a central role in the study of strongly correlated systems. Their sensitivity to vacuum degeneracy, symmetry breaking, and spectral flow makes them ideal tools for exploring features of the $\theta$-vacuum.

Quantum simulation, exact diagonalization and tensor network methods now allow for controlled, nonperturbative studies of these observables in lattice gauge theories. In particular, several studies have applied entanglement and topological diagnostics to chart the phase diagram of the massive Schwinger model. For example, matrix product states (MPS) calculations \cite{Buyens:2017crb,LenaPRD2020} confirmed a first-order transition at large $m/g$ and investigated the $\theta$ dependence of observables such as the chiral condensate in the continuum limit (restricted to small $\theta$). 

The lattice Hamiltonian formulation of the Schwinger model enables the use of tensor-network methods and exact diagonalization to compute the reduced density matrix of a subsystem $A$ and extract its entanglement spectrum.  However, reconstructing ES through full state tomography becomes exponentially inefficient with system size and it is not directly accessible experimentally.

Recent developments based on the Bisognano--Wichmann (BW) theorem suggest a conceptual shift in how entanglement properties may be accessed. Instead of reconstructing the density matrix $\rho_A$, one may directly probe the corresponding entanglement (or modular) Hamiltonian (EH), defined through $\rho_A = e^{-\bar{H}_A}$. In relativistic quantum field theory, the BW theorem provides an exact and local expression for the modular Hamiltonian of a half-space bipartition. Building on this, Ref. \cite{Dalmonte_2018} reformulated the BW construction for finite lattice systems and demonstrated that a lattice Bisognano--Wichmann (LBW) ansatz accurately reproduces the low-lying entanglement spectrum in a wide range of strongly correlated models. These include conformal phases of spin and fermionic systems, topological phases in one and two dimensions, and quantum critical regimes of Ising-type models \cite{Peschel,Peschel_2009,PhysRevLett.58.1395,Eisler_2017,Eisler_2017}. In all cases, the agreement is controlled by the emergence of approximate Lorentz invariance at low energies.

In this work, we apply the LBW construction to the lattice formulation of the massive Schwinger model. 
{Since Lorentz invariance is explicitly broken at finite lattice spacing, the BW theorem does not apply exactly to the lattice theory. 
We therefore investigate whether the low-energy sector of the lattice model is sufficiently close to the relativistic continuum regime for the LBW ansatz to accurately reproduce the exact modular Hamiltonian obtained from the reduced density matrix.
To assess this, we compare the corresponding eigenvectors obtained from the LBW ansatz with those of the exact modular Hamiltonian. Agreement between the two would indicate that the infrared physics of the lattice model is well described by the emergent BW structure expected from the continuum theory.
Such a correspondence implies that the modular Hamiltonian can be effectively interpreted as a spatially weighted version of the physical Hamiltonian. This provides a direct connection between entanglement properties and the underlying low-energy excitation spectrum, allowing entanglement features to be related to physical quantities such as the excitation gap.}
Moreover, establishing the validity of the LBW description opens the possibility of simulating the entanglement Hamiltonian directly on quantum hardware, such as IBM quantum processors, without requiring full reconstruction of the reduced density matrix $\rho_A$. 

\bigskip

The paper is organized as follows. 
Section II introduces the massive Schwinger model with a $\theta$ term, its chirally rotated lattice formulation, the lattice corrections, and the observables used in our analysis.
Section III presents the $\theta$ and mass dependence of the vacuum energy density, chiral condensate, electric field, and entanglement observables, together with the topological susceptibility and spatial correlation functions. It also tests the lattice Bisognano--Wichmann ansatz through a comparison of entanglement eigenvectors.
Section IV summarizes our findings and discusses their implications and possible extensions.

\section{Model and methods}
\label{sec:theory}
\subsection{Schwinger Model with $\theta$-term}
The massive Schwinger model describes (1+1)-dimensional quantum electrodynamics with a single Dirac fermion $\psi$ coupled to a $U(1)$ gauge field $A_\mu$. 
The action in Minkowski space is defined by 
\begin{equation}
    \mathcal{S} = \int d^2x \left[ -\frac{1}{4}F_{\mu\nu}F^{\mu\nu} + \frac{g \theta }{4\pi}\epsilon_{\mu\nu} F^{\mu\nu} + \bar{\psi}(i\slashed{D} - m)\psi \right],
\end{equation}
after a chiral rotation $\psi\rightarrow e^{i\gamma_5\theta/2}\psi$ and $\bar{\psi}\rightarrow\bar{\psi}e^{i\gamma_5\theta/2}$,
\begin{equation}
    \mathcal{S} = \int d^2x \left[ -\frac{1}{4}F_{\mu\nu}F^{\mu\nu} +  \bar{\psi}i\slashed{D} \psi - m\bar{\psi} e^{i \gamma_{5} \theta}\psi  \right]
\end{equation}
\noindent where $\slashed{D}=\gamma^\mu (\partial_\mu + ig A_\mu)$, $m$ is the bare fermion mass, and $g$ is the gauge coupling constant. In 1+1 dimensions, the field strength tensor $F_{\mu\nu} = \partial_\mu A_\nu - \partial_\nu A_\mu$ reduces to a single component, $F_{01}=E$, {corresponding to an electric field. Unlike higher dimensions, there are no propagating photon degrees of freedom, and the electric field is completely determined by the fermionic charge and current distributions through Maxwell’s equations.} The action includes a topological $\theta$-term, a total derivative that does not affect the classical equations of motion. While classically $\theta$ can be any real value, in the quantum theory it introduces an additional phase in the path integral and leads to nontrivial vacuum structure and CP violation. Moreover, $\theta$ becomes a periodic variable with $\theta \in [0, 2\pi)$, so only values modulo $2\pi$ are physically meaningful \cite{PhysRevLett.42.1195}.

This fermionic gauge theory admits an equivalent bosonized formulation in terms of a single real scalar field $\phi$. The bosonized action reads
\begin{equation}
\mathcal{S}[\phi] = \int d^2x\, \left[ -\frac{1}{2} (\partial_\mu \phi)^2 - V(\phi) \right],
\end{equation}
with the effective potential {given by \cite{Coleman1976, Grieninger_2024}}
\begin{equation}
V(\phi) = \frac{1}{2} \mu^2 \phi^2 - cm\mu \cos(2\sqrt{\pi} \phi + \theta).
\label{boso_pot}
\end{equation}
Here, $\mu = g/\sqrt{\pi}$ sets the mass of the scalar boson in the massless Schwinger model, and $c=e^\gamma /2\pi$ is a dimensionless coefficient with Euler constant $\gamma \approx 0.577$. 

The potential $V(\phi)$ consists of a confining quadratic term and a periodic cosine term that is proportional to the fermion mass. To characterize the relative strength of the two competing terms, we define the dimensionless parameter \cite{Batini_2024}
\begin{equation}
\kappa = 2\sqrt{\pi} \, \frac{cm\mu }{\mu^2} = \frac{e^\gamma m }{g},
\label{kappa_term}
\end{equation}
which provides an alternative way to express the ratio $m/g$. 

The minima of the potential $V(\phi)$ define the vacuum configurations, and their locations depend on the value of the topological angle $\theta$. In the strong coupling limit, where $g^2 \gg m^2$ or equivalently $\kappa \ll 1$, the quadratic term dominates, and the potential supports a single, well defined vacuum. Conversely, in the weak coupling regime $m \gg g$ or $\kappa \gg 1$, the cosine term becomes dominant, leading to the emergence of multiple, nearly degenerate minima.

\begin{figure}[t!]
    \centering
     \begin{subfigure}
         \centering
         \includegraphics[width=0.9\linewidth]{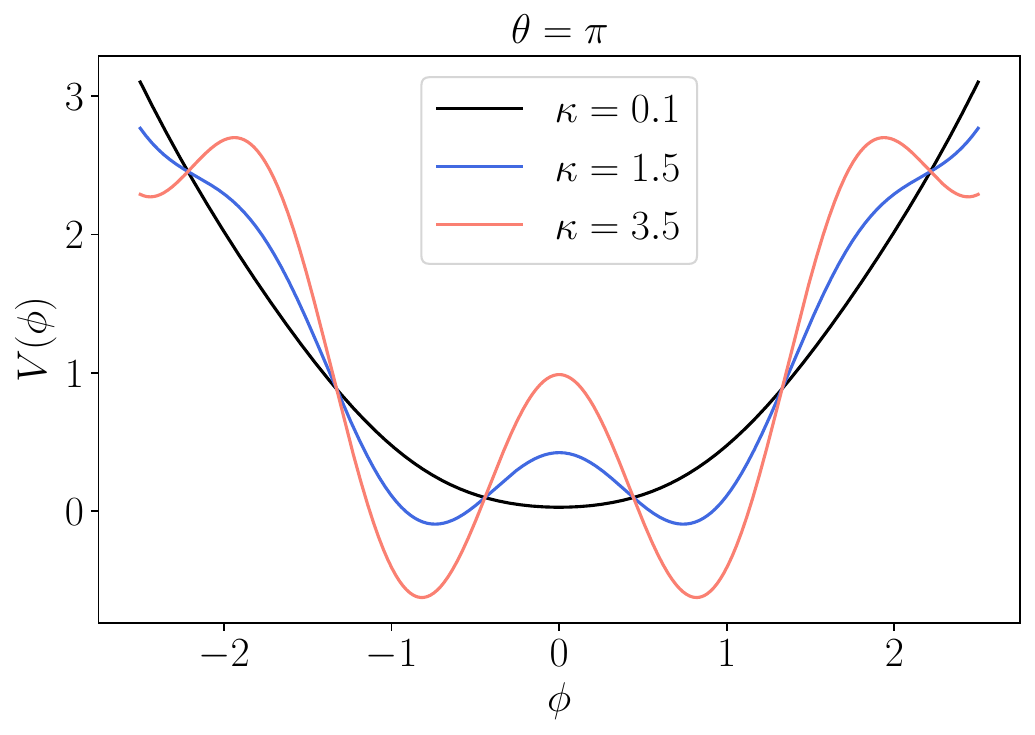}
     \end{subfigure}
     \hfill
    \caption{Potential $V(\phi)$ of the bosonized massive Schwinger model for different values of the dimensionless parameter $\kappa=e^\gamma m/g$, for fixed $\theta=\pi$, where CP is preserved and degenerate vacua appear in the weak-coupling limit.}
    \label{potkappa_theta}
\end{figure}

Fig.~\ref{potkappa_theta} shows the effective potential $V(\phi)$ for different values of $\kappa$ and fixed $\theta$. For small $\kappa$, the potential is dominated by the quadratic term, resulting in a single minimum near $\phi = 0$. As $\kappa$ increases, the periodic term introduces additional vacuum structure, leading to the appearance of multiple local minima. These structures depend on the value of $\theta$. 
For $\theta = 0$ or $\pi$, the potential remains symmetric under $\phi \to -\phi$, preserving CP symmetry.

\subsection{Implementation of the $\theta$ term on the lattice}

We now clarify the motivation behind our choice of lattice formulation for the massive Schwinger model with a topological $\theta$ term, and its relation to earlier continuum arguments due to Witten \cite{Witten:1980sp, Witten_1998} regarding the origin of $\theta$ periodicity.

In the continuum, the angle $\theta$ is a compact parameter, and all physical observables must be $2\pi$ periodic in $\theta$. However, this periodicity is not generally manifested at the level of the Lagrangian or Hamiltonian. As emphasized by Witten in his analysis of the chiral dynamics and large $N$ gauge theories \cite{Witten:1980sp}, the $\theta$ dependence of the theory is intrinsically multi-branched: the effective potential for the scalar (or pseudoscalar) field consists of several branches, each of which is individually non-periodic in $\theta$. The physical $2\pi$ periodicity emerges only after minimizing the vacuum energy over these branches, corresponding to distinct topological sectors.
Then, $\theta$ periodicity is an emergent property of the vacuum selection, rather than a property of the Hamiltonian.

These considerations have direct implications for lattice implementations of the $\theta$ term. A common approach introduces $\theta$ by shifting the electric-field operator $L_n$ at site $n$ {by $\theta/(2\pi)$}, so that the gauge-field energy acquires a quadratic dependence on $\theta$. 
The resulting Hamiltonian is not periodic in $\theta$, since values differing by $2\pi$ correspond to inequivalent Hamiltonians. In this formulation the lattice theory itself does not enforce the compact nature of $\theta$.

In the continuum Schwinger model, the expected $2\pi$ periodicity arises only after minimizing the vacuum energy over all electric-flux sectors, in close analogy with the vacuum-selection mechanism discussed in \cite{Witten:1980sp}. This procedure selects the lowest-energy branch among the different flux sectors and thereby establishes the periodic dependence on $\theta$. On the lattice, however, this vacuum-selection mechanism is not fully realized, and the Hamiltonian may therefore retain a non-periodic dependence on $\theta$.

An alternative implementation avoids this issue by performing a chiral rotation of the fermion fields.  Due to chiral anomaly, this transformation transfers the $\theta$ dependence from the gauge sector to the mass term. The resulting lattice Hamiltonian contains explicit trigonometric $\theta$ dependence multiplying fermion bilinears, and is therefore manifestly $2\pi$ periodic in $\theta$ already at the operator level.

This distinction becomes particularly important in the massless limit. In the continuum theory, $\theta$ becomes unphysical at $m=0$ because chiral rotation can remove it, so gauge-invariant observables are $\theta$-independent. A lattice formulation that implements $\theta$ as a background electric field can nevertheless exhibit residual $\theta$ dependence at the Hamiltonian level prior to vacuum selection. In studies employing this background-field formulation, Ref. \cite{LenaPRD2020} observes that even in the massless limit a $\theta$-dependent lattice artifact is observed, and the expected behavior is recovered only after taking the continuum limit.

Finally, we note that the distinction between these formulations is relevant primarily for {open boundary conditions (OBC)}. For periodic boundary conditions, physical observables are $2\pi$ periodic in $\theta$ irrespective of the chosen lattice implementation, as the summation over topological sectors is effectively enforced by the boundary conditions \cite{dempsey2023discretechiralsymmetrymass}. In contrast, for OBC, used throughout this work, the realization of $\theta$ periodicity depends sensitively on how the $\theta$ term is implemented at the Hamiltonian level.

For these reasons, while both formulations reproduce the same continuum physics after the appropriate vacuum selection, they differ conceptually and practically at finite lattice spacing. Since our goal is to study the $\theta$ dependence across its full range while preserving the correct massless limit and making the compact nature of $\theta$ explicit at the operator level, we adopt the chirally rotated formulation throughout this work.

\subsection{Lattice Hamiltonian formulation}
We discretize the Schwinger model on a spatial lattice in the Hamiltonian formalism and work in temporal gauge, $A_0=0$. This formulation is convenient for numerical simulations and allows a direct mapping to spin degrees of freedom.
We use staggered fermions, which reduce fermion doubling while preserving a remnant of the continuum chiral symmetry \cite{PhysRevD.11.395,PhysRevD.16.3031}.

In temporal gauge,~\footnote{See \cite{Yao:2025uxz} for a discussion of different gauges.} the canonical momentum conjugate to $A_1$ is the electric field, $E=\dot A_1$, and the continuum Hamiltonian takes the form
\begin{equation}
H=\int dx\left[\frac12 E^2 - i\bar\psi\gamma^1(\partial_1+igA_1)\psi
+ m\,\bar\psi e^{i\theta\gamma_5}\psi\right].
\end{equation}
The $\theta$ dependence has been moved into the fermion mass term via chiral rotation, using the chiral anomaly. This representation is especially useful on the lattice, since the $\theta$ dependence appears explicitly periodic in fermionic bilinears, while the electric-field term retains its standard quadratic form.

\subsubsection{Staggered lattice discretization}

We discretize space as $z=an$, where $a$ is the lattice spacing and $n\in\mathbb{Z}$ labels the lattice sites. 
In the staggered formulation, the two components of the Dirac spinor $\psi=(\psi^1,\psi^2)^T$ are encoded in a single-component lattice field $\chi_n$ according to
\begin{equation}
\psi^1 \rightarrow \frac{\chi_n}{\sqrt{a}} \quad \text{for odd } n, 
\qquad 
\psi^2 \rightarrow \frac{\chi_n}{\sqrt{a}} \quad \text{for even } n.
\end{equation}

The gauge and electric-field operators live on the links between neighboring sites and are defined by
\begin{equation}
L_n = \frac{E(an)}{g} , \qquad U_n = e^{-iag A_1(an)},
\end{equation}
where $U_n$ is the parallel transporter from site $n$ to site $n+1$.

With these definitions, the lattice Hamiltonian ($H^L_S$) in terms of the staggered fermion field $\chi_n$ is
\begin{align}
&H^L_S = H^L_{F} + H^L_{E}, \nonumber \\
&H^L_{F} = \sum_{n=1}^{N} m (-1)^n \cos \theta\, \chi_n^\dagger \chi_n  \nonumber \\
& -\sum_{n=1}^{N-1} \left( \frac{im}{2} (-1)^n \sin \theta + \frac{i}{2a} \right) 
\left( U_n \chi_{n+1}^\dagger \chi_n - U_n^\dagger \chi_n^\dagger \chi_{n+1} \right) \nonumber \\
&H^L_E= \frac{ag^2}{2} \sum_{n=1}^{N-1} L_n^2,
\label{HamiltonianChi}
\end{align}
where $N$ is taken to be even. 
The first term contains the fermionic kinetic hopping, the staggered mass term, and the $\theta$-dependent pseudoscalar contribution generated by the chiral rotation. 
The second term represents the electric-field energy.

\subsubsection{Gauss's law}
{We seek the bulk vacuum on the infinite line and therefore use a large finite interval with fixed boundary electric flux. Periodic boundary conditions would instead describe a spatial circle, where Gauss's law leaves a global electric-flux mode and its conjugate Wilson-loop holonomy that must be retained. OBC also allow the link variables to be eliminated using Gauss's law, as shown below. Because a finite interval has boundaries, we interpret our results as bulk quantities only after checking that the central region is insensitive to them.}

{With OBC}, and with $\chi_0=\chi_{N+1}=0$ and a fixed boundary electric field, the link variables $U_n$ can be removed by a local gauge transformation,

\begin{equation}
\chi_n \rightarrow g_n \chi_n, \qquad U_n \rightarrow g_{n+1} U_n g_n^\dagger,
\end{equation}
with $g_1=1$ and $g_n=\prod_{m=1}^{n-1}U_m^\dagger$. 
This gauge fixing eliminates the explicit link operators from the Hamiltonian, while the electric fields remain constrained by Gauss's law. The continuum relation $\partial_z E = g\bar\psi\gamma^0\psi$ becomes on the lattice
\begin{equation}
L_n - L_{n-1} = Q_n,
\end{equation}
where the local charge operator is
\begin{equation}
Q_n = \chi_n^\dagger \chi_n - \frac{1 - (-1)^n}{2}.
\label{eq:charge}
\end{equation}
Fixing the total charge sector and setting $L_0=0$, one obtains
\begin{equation}
L_n = \sum_{m=1}^n Q_m.
\end{equation}
As usual in the Schwinger model, solving Gauss's law in this way trades the gauge links for a nonlocal Coulomb interaction encoded in the electric-field energy.

\subsubsection{Discrete chiral symmetry}
A useful feature of the continuum massless Schwinger model is that the Hamiltonians at $\theta$ and $\theta+\pi$ are related by a discrete chiral transformation. 
In the staggered lattice formulation, the corresponding operation is realized by translation by one lattice site.

In the periodic lattice theory, where one-site translations are implemented by a unitary operator, let $\mathcal V$ be defined by
\begin{equation}
\mathcal V\chi_n \mathcal V^{-1}=\chi_{n+1},
\qquad
\mathcal VU_n \mathcal V^{-1}=U_{n+1}.
\end{equation}
Because the staggered factor changes sign under a one-site shift, the scalar density transforms as
\begin{equation}
\mathcal V\sum_n(-1)^n\chi_n^\dagger\chi_n \mathcal V^{-1}
=
-\sum_n(-1)^n\chi_n^\dagger\chi_n,
\end{equation}
and the same sign change occurs for the pseudoscalar hopping bilinear. 
It follows that the fermionic part of the Hamiltonian in Eq. \eqref{HamiltonianChi} obeys 
\begin{equation}
\mathcal V H^L_{F}(\theta)\mathcal V^{-1}= H^L_{F}(\theta+\pi).
\end{equation}

{On the staggered lattice, the translation operator affects charge as ${\cal V} Q_n {\cal V}^{-1}=Q_{n+1}+(-1)^n$ and Gauss's law fixes its action on the flux,
\begin{equation}
\mathcal V L_n\mathcal V^{-1}=L_{n+1}+\frac{1+(-1)^n}{2}+l ,
\label{Lshift}
\end{equation}
where $l$ is an integer~\cite{dempsey2023discretechiralsymmetrymass}.
In general, an electric term with a uniform background field
\begin{equation}
    H_E^L(\alpha)=\frac{ag^2}{2}\sum\limits_n (L_n+\alpha)^2
\end{equation}
transforms nontrivially because Gauss's law induces a shift in the electric field under translation. 
Using the lattice Gauss-law constraint, one finds
\begin{equation}
\mathcal V H_E^L(\alpha) \mathcal V^{-1}
=
H_E^L(\alpha+l+\frac12)
-\frac{ag^2}{4}\sum_n(-1)^n\chi_n^\dagger\chi_n .
\end{equation}}
Combining the fermionic and gauge contributions then gives {(up to the background shift)}
\begin{equation}
\mathcal V H^L_S(\theta)\mathcal V^{-1}
=
H^L_S(\theta+\pi)
-\frac{ag^2}{4}\sum_n(-1)^n\chi_n^\dagger\chi_n .
\end{equation}

Thus, translation by one site does not map the lattice Hamiltonian exactly to $H^L_{S}(\theta+\pi)$. 
The mismatch is {(up to the background shift)} proportional to the staggered scalar density and therefore represents a finite-lattice artifact that explicitly breaks the discrete chiral symmetry.

This artifact can be canceled by adding the counter term
\begin{equation}
\delta H =\Delta \sum_n(-1)^n\chi_n^\dagger\chi_n ,
\end{equation}
which changes sign under the one-site translation. 
Demanding exact covariance under $\theta\to\theta+\pi$ fixes
\begin{equation}
\Delta =-\frac{ag^2}{8}.
\end{equation}
The improved lattice Hamiltonian is therefore
\begin{equation}
H^L
=
H^L_S
-
\frac{ag^2}{8}\sum_n(-1)^n\chi_n^\dagger\chi_n .
\end{equation}

The derivation above assumes the periodic lattice theory, where the one-site translation operator is well defined. In the open-boundary simulations considered in this work, the boundaries explicitly break this translation symmetry. Consequently, the correction term should not be interpreted as restoring an exact symmetry of the finite OBC Hamiltonian. Rather, it acts as a bulk improvement term that removes the leading lattice artifact associated with the staggered realization of the discrete chiral transformation.

This correction vanishes in the continuum limit $a\to0$. At finite lattice spacing it restores the discrete chiral symmetry of the periodic lattice theory and improves the approach to the continuum limit in both periodic and open-boundary calculations.
By applying the discrete chiral symmetry transformation to the chirally rotated Hamiltonian, we find the same lattice correction $\delta H$ previously identified for the standard lattice Hamiltonian in Ref. \cite{dempsey2023discretechiralsymmetrymass}. 
In the chirally rotated formulation, this correction is an additive shift of the staggered scalar coefficient. At $\theta=0$, it gives the coefficient $m-ag^2/8$. For general $\theta$, however, it cannot be implemented by replacing $m$ with $m-ag^2/8$ in both rotated mass terms, since the pseudoscalar coefficient $m\sin\theta$ remains unchanged.

The improved Hamiltonian is used throughout this work and leads to improved scaling of the lattice results toward the continuum limit.

\subsubsection{Spin representation}
 Choosing to work in the basis where the Dirac matrices are $\gamma_0= \sigma_z \equiv Z$, $\gamma_1= -i\sigma_y \equiv  -iY$, and $\gamma_5={-}\gamma_0\gamma_1 ={-}\sigma_x \equiv -X$, we map the fermionic theory to a spin chain by applying the Jordan-Wigner transformation \cite{Jordan:1928wi}
\begin{align}
\chi_n &= \frac{X_n - iY_n}{2} \prod_{j=1}^{n-1} (-iZ_j), \nonumber \\
\chi_n^\dagger &= \frac{X_n + iY_n}{2} \prod_{j=1}^{n-1} (iZ_j).
\end{align}
The Hamiltonian used in the numerical calculations is therefore given below, together with the Gauss-law solution
\begin{align}
H^L = & \sum_{n=1}^{N-1} \left( \frac{1}{4a} + \frac{m}{4} (-1)^n \sin \theta \right) 
\left( X_n X_{n+1} + Y_n Y_{n+1} \right) \nonumber \\
& + \frac{1}{2}\sum_{n=1}^{N} \left( m \cos \theta - \frac{ag^2}{8} \right)(-1)^n Z_n \nonumber \\
&+ \frac{ag^2}{2} \sum_{n=1}^{N-1} L_n^2,
\label{finalHam}
\end{align}
with
\begin{equation}
L_n = \sum_{j=1}^{n} \frac{Z_j +(-1)^j}{2}.
\end{equation}
The first term describes nearest-neighbor hopping written as an XY exchange interaction, the second is the staggered mass contribution including the symmetry-restoring counterterm, and the last term encodes the long-range Coulomb interaction generated by Gauss's law.

\subsection{Chirally rotated expectation values}
\label{chirally_rot_obs}
\paragraph{Chiral condensate.}
Under a chiral transformation the scalar condensate transforms as
\begin{equation}
\bar{\psi}\psi 
\;\xrightarrow{U(1)_R}\;
\bar{\psi} e^{i\theta\gamma_5}\psi
=
\cos\theta\,\bar{\psi}\psi
+
i\sin\theta\,\bar{\psi}\gamma_5\psi .
\end{equation}

Using the staggered-fermion dictionary summarized in Ref.~\cite{Kharzeev_2020}, the corresponding lattice operator for the rotated condensate can be written as
\begin{align}
\braket{\bar{\psi}\psi}_{R}&= \cos{\theta} \braket{\bar{\psi}\psi} + \sin{\theta} \braket{\bar{\psi}i\gamma_5\psi}   \nonumber \\
& =
\cos\theta\, \Bigg\langle \sum_n \frac{(-1)^n}{2a} Z_n \Bigg\rangle
\nonumber\\&\quad+\sin\theta\,\Bigg\langle \sum_n \frac{(-1)^n}{4a} 
\left( X_n X_{n+1}+Y_n Y_{n+1} \right) \Bigg\rangle.
\label{Eq:chiralcondensate}
\end{align}
This operator is manifestly CP-even, the scalar contribution is weighted by $\cos\theta$ (both CP-even), while the hopping term is weighted by $\sin\theta$ (both CP-odd), so that each term is CP-even and the total combination remains CP-even.

\paragraph{Electric field.}

The electric field can be related to fermionic observables through the anomalous Ward identity,
\begin{equation}
\partial_\mu j_5^\mu
=
-2im\,\bar{\psi}\gamma_5\psi
+
\frac{g}{\pi}E,
\end{equation}
where the chiral current is \(j_5^\mu=\bar{\psi}\gamma^\mu\gamma^5\psi\).

Upon averaging over the lattice, the derivative term becomes a total difference and therefore vanishes (exactly for periodic boundary conditions and up to boundary corrections for open chains). 
The spatially averaged electric field then satisfies the identity
\begin{align}
{E}_{WI} &= \frac{2\pi m}{g} \langle \bar{\psi} i\gamma_5 \psi \rangle \label{eq:EWI} \\
&= \frac{2\pi m}{g} \left[ \cos{\theta}\braket{\bar{\psi} i\gamma_5 \psi} - \sin{\theta}\braket{\bar{\psi}\psi}\right]
\end{align}

Using the spin representation of the pseudoscalar condensate, this relation can be written as
\begin{align}
{E}_{WI}  =& \frac{2\pi m}{g}\,
\Bigg\langle \frac{1}{N} \Bigg(\!\! \cos{\theta} \sum_n \frac{(-1)^n}{4a}
\!\left( X_n X_{n+1}+Y_n Y_{n+1} \right) \nonumber \\
&- \sin{\theta} \sum_n \frac{(-1)^n}{2a} Z_n \Bigg)\Bigg\rangle .
\label{Eq:avgEfield}
\end{align}

This identity shows that the spatially averaged electric field is related, through the anomalous Ward identity, to the \emph{CP-odd} fermion bilinear in the chirally rotated basis. 
In contrast, the rotated chiral condensate in Eq.~\eqref{Eq:chiralcondensate} corresponds to a \emph{CP-even} combination. 

This distinction can be seen explicitly from the structure of the rotated operators. In the condensate, the scalar contribution is weighted by $\cos\theta$ (both CP-even), while the hopping term is weighted by $\sin\theta$ (both CP-odd), so that each term is CP-even and the total operator is CP-even. In contrast, for the electric field the coefficients are interchanged: the pseudoscalar contribution appears with $\cos\theta$ and the scalar one with $\sin\theta$, yielding terms of the form even$\times$odd and odd$\times$even, which are both CP-odd. The electric field therefore probes the CP-odd combination of fermion bilinears. Note also that $E_\text{WI}$ and the chiral condensate are related in the following way: integrating the chiral condensate expression with respect to m and taking a derivative with respect to $\theta$ (weighted by 2$\pi/g$) yields the same expression as $E_\text{WI}$. This is exactly the same relation as is used in chiral perturbation theory where the chiral condensate follows as a mass derivative of $E_0$ and the electric field is proportional to the $\theta$ derivative of $E_0$ (see eq.~\eqref{eq:condpert} and eq.~\eqref{eq:elfield}).

\subsection{Lattice contact term of chirally rotated Hamiltonian}
\label{contact_terms}

{Discretizing the chirally rotated staggered Hamiltonian introduces a $\theta$ dependent contribution to the vacuum energy density that does not vanish in the continuum limit and thus has to be subtracted. In this section, we derive the contact term by matching the rotated lattice vacuum energy to the continuum theory.}

{In the following, we work with the free theory $(g=0)$ where a constant chiral rotation of a free
Dirac mass does not change the vacuum energy. This allows us to obtain the counter term in closed form.}

{The $\cos\theta$ term in Eq.~\eqref{finalHam} (with $g=0$) is the staggered site mass term
$\mu=m\cos\theta$, while the $\sin\theta$ term alternates the nearest-neighbor hopping,}
{\begin{equation}
t_n=\frac{1}{2a}+(-1)^n\frac{m}{2}\sin\theta.
\end{equation}}

{Since one physical site corresponds to two staggered sites, the single-particle dispersion relation is}
{\begin{equation}
E_\theta(k)^2
=
m^2+
\left(\frac{1}{a^2}-m^2\sin^2\theta\right)\cos^2(ka),
\qquad
|k|\leq\frac{\pi}{2a},
\label{eq:rotdisp}
\end{equation}}
{which is exact at arbitrary $ma$ and $\theta$. Note that the additional contribution vanishes at $\theta=0$ and integer multiples of $\pi$.}

{In the thermodynamic limit, the excess of the energy density at half-filling introduced by the $\theta$ dependent term is captured by}
{\begin{equation}
c_a^{(0)}(\theta)
\equiv
\mathcal E_a^{(0)}(\theta)-\mathcal E_a^{(0)}(0)
=
\frac{1}{2\pi}
\int_{-\pi/2a}^{\pi/2a} dk\,
\bigl[E_0(k)-E_\theta(k)\bigr].
\label{eq:contactexact}
\end{equation}}

{Rescaling $u=ka$ and writing $\epsilon=ma$ puts the whole $\theta$ dependence into the single factor $\lambda=1-\epsilon^2\sin^2\theta$ multiplying $\cos^2u$,}
{\begin{align}
c_a^{(0)}(\theta)=\frac{1}{2\pi a^2}\int_{-\pi/2}^{\pi/2}\!\!du\,
\Bigl[&\sqrt{\epsilon^2+\cos^2u} 
-\sqrt{\epsilon^2+\lambda\cos^2u}\,\Bigr].
\label{eq:contactrescaled}
\end{align}}

{For $|ma|\ll1$, the excess reads
\begin{equation}
c_a^{(0)}(\theta)
=
\frac{m^2\sin^2\theta}{2\pi}
+
O\!\left(
m^4a^2\ln\frac{1}{|ma|}
\right).
\label{eq:contactasympt}
\end{equation}}

{In the continuum limit, it is sufficient to subtract the free counter term  $c_a^{(0)}(\theta)$ from the observables. Following the normalizations of \cite{LenaPRD2020}, we subtract
\begin{align}
\left.
\frac{\Delta\mathcal E_0}{g^2}
\right|_{\rm ct}
&=
\frac{c_a^{(0)}(\theta)}{g^2},
&
\left.
\frac{\Delta\Sigma}{g}
\right|_{\rm ct}
&=
\frac{1}{g}
\frac{\partial c_a^{(0)}(\theta)}{\partial m},
\nonumber\\
\left.
\frac{E_{\rm WI}}{g}
\right|_{\rm ct}
&=
\frac{2\pi}{g^2}
\frac{\partial c_a^{(0)}(\theta)}{\partial\theta},
&
\left.
\frac{\chi_{\rm top}}{g^2}
\right|_{\rm ct}
&=
\frac{1}{g^2}
\left.
\frac{\partial^2 c_a^{(0)}(\theta)}
     {\partial\theta^2}
\right|_{\theta=0},
\label{eq:subtractions_exact}
\end{align}}
{where $E_{\rm WI}$ is defined in eq.~\eqref{eq:EWI}, and $\chi_{\mathrm{top}}$ is the topological susceptibility.
These are exact finite-$a$ derivatives of the free-lattice subtraction used here.
They do not include the unknown interaction-dependent matching at nonzero $g$ (which we found to be of order $(ga)^2$). Note that while $E_{\rm WI}$ needs the counter term subtraction, $E$ computed directly from $\langle \psi^\dagger \psi\rangle$ does not. With the counter term subtraction, both ways of computing the electric field agree. Without it, there is strong disagreement at finite $\theta$ for larger $m/g$. }

\subsection{Entanglement entropy and spectrum}
We characterize the quantum correlations of the ground state using bipartite entanglement. In our numerical simulations, the one-dimensional lattice is partitioned at its midpoint into left and right subsystems, corresponding to a factorization $\mathcal{H}=\mathcal{H}_A\otimes\mathcal{H}_B$ of the Hilbert space.
Starting from the ground state $|\Psi_0\rangle$ obtained numerically, we construct the reduced density matrix of the left subsystem,
\begin{equation}
    \rho_A = {\rm Tr}_B
    \left(|\Psi_0\rangle\langle\Psi_0|\right).
\end{equation}
Its spectral decomposition is
\begin{equation}
    \rho_A
    =
    \sum_{i=1}^{\chi}
    \lambda_i
    |\psi_i^A\rangle\langle\psi_i^A|,
\end{equation}
where $\lambda_i\geq0$ are the eigenvalues of $\rho_A$, normalized as $\sum_i\lambda_i=1$, and $\chi$ is the Schmidt rank.

The set of eigenvalues $\{\lambda_i\}$ of the reduced density matrix, which we refer to as the Schmidt eigenvalues, constitute the entanglement
spectrum. For visualization, we represent the spectrum in terms of $\log_{10}\lambda_i$, which makes the separation and reorganization of subleading entanglement levels more apparent.

{We resolve the reduced density matrix into sectors of the subsystem charge \(Q_A=\sum_{n\in A}Q_n\), with \(Q_n\) defined in Eq. \eqref{eq:charge}. For a ground state with fixed total charge, \([\rho_A,Q_A]=0\), so that
\begin{equation}
\rho_A=\bigoplus_q\rho_A^{(q)}.
\end{equation}
We denote the eigenvalues of these blocks by \(\lambda_{q,\alpha}\), where \(q\) labels the subsystem charge and \(\alpha\) labels the levels within each sector. These are weights of the full reduced density matrix, normalized by \(\sum_{q,\alpha}\lambda_{q,\alpha}=1\), rather than eigenvalues of separately normalized charge blocks. We refer to this charge-labeled decomposition as the symmetry-resolved entanglement spectrum.}

The von Neumann entanglement entropy is given by
\begin{equation}
    S_{\rm EE} =
    -{\rm Tr}\left(\rho_A\log\rho_A\right) =
    -\sum_i\lambda_i\log\lambda_i,
\end{equation}
and captures the total amount of bipartite entanglement across the cut. In contrast, the full entanglement spectrum contains additional information, such as level degeneracies or characteristic splittings, which may signal emergent low-energy degrees of freedom, competing vacuum configurations, or changes in the underlying vacuum structure.
{We track the charge-resolved Schmidt weights as functions of \(\theta\) and \(m/g\). The symmetry resolved spectrum allows to distinguish degeneracies between opposite-charge sectors from the broader compression of the low-lying entanglement spectrum near the critical region.}

{The entanglement entropy of a spatial region is ultraviolet divergent in continuum quantum field theory due to short-distance correlations across the entangling surface~\cite{Casini_2009}. In the lattice formulation, the lattice spacing $a$ provides a natural UV regulator. The leading UV contribution originates from local short-distance correlations and is expected to be largely independent of the vacuum angle $\theta$. For this reason, we focus on the variation of the entanglement entropy relative to its value at $\theta=0$, $ \Delta S_{\rm EE}(\theta) = S_{\rm EE}(\theta)-S_{\rm EE}(0)$, for which the leading cutoff-dependent contribution largely cancels.
The $\theta$ dependence of $\Delta S_{\rm EE}$ and of the entanglement spectrum can therefore be used to probe changes in the infrared vacuum structure of the Schwinger model.}

In practice, we diagonalize $\rho_A$ for a bipartition at the center of the chain and track the evolution of both the entanglement entropy and the low-lying entanglement spectrum as functions of the topological angle $\theta$ and the mass-to-coupling ratio $m/g$.

\subsection{Entanglement Hamiltonians and the Bisognano--Wichmann theorem}
An alternative way to characterize the structure of quantum entanglement is to express the reduced density matrix in terms of an effective operator acting only within the subsystem. For any bipartition $A \cup B$, the reduced density matrix of region $A$ can always be written in exponential form,
\begin{equation}
\rho_A = e^{-\bar H_A},
\end{equation}
thereby defining the entanglement (or modular) Hamiltonian $\bar H_A$. The spectrum of $\bar H_A$, known as the entanglement spectrum, provides a characterization of quantum correlations beyond entanglement entropy.

In general, the modular Hamiltonian $\bar H_A$ is a highly nonlocal operator and does not commute with the physical Hamiltonian $H$ of the full system $A \cup B$,
\begin{equation}
[H,\bar H_A] \neq 0,
\end{equation}
reflecting the fact that $H$ generates real-time evolution, whereas $\bar H_A$ generates modular flow associated with the reduced density matrix.

A remarkable simplification occurs in relativistic quantum field theories through the Bisognano--Wichmann (BW) theorem. For a Lorentz-invariant vacuum state and a spatial bipartition defined by a half-space $A=\{x_1>0\}$, the BW theorem provides an exact expression for the modular Hamiltonian,
\begin{equation}
\bar H_A = 2\pi \int_{x_1>0} d^d x \; x_1 \, \mathcal{H}(x) + c_0,
\end{equation}
where $\mathcal{H}(x)$ is the local Hamiltonian density and $c_0$ ensures normalization of $\rho_A$. This expression identifies the entanglement Hamiltonian with the generator of Lorentz boosts restricted to region $A$.

The origin of this result lies in spacetime symmetries. The Hamiltonian $H$ and the boost generator satisfy the Poincaré algebra,
\begin{equation}
[H,K] = iP,
\end{equation}
where $P$ is the momentum operator. For a Lorentz invariant vacuum, translational invariance implies $P|\Psi_0\rangle = 0$, so the vacuum is invariant under both time translations and boosts. 

Since $\bar H_A$ and $\rho_A$ are related by an exponential map, they are simultaneously diagonalizable by construction. Their eigenvectors therefore coincide, while their eigenvalues are related by
\begin{equation}
\lambda_\alpha = e^{-\epsilon_\alpha},
\end{equation}
where $\lambda_\alpha$ are the eigenvalues of $\rho_A$ and
$\epsilon_\alpha$ are the corresponding eigenvalues of $\bar H_A$. This establishes a direct correspondence between the entanglement spectrum and the spectrum of the modular Hamiltonian.

On the lattice, Lorentz invariance is explicitly broken, and the BW theorem no longer holds exactly. In particular, the exact modular Hamiltonian $\bar H_A = -\log \rho_A$ is generically a highly nonlocal many-body operator. Nevertheless, if the low-energy physics of the lattice model flows to a relativistic continuum theory, one may expect the BW structure to hold approximately in the infrared.
Evidence for such an emergent Bisognano--Wichmann structure has been observed in several lattice settings \cite{Dalmonte_2018, Giudici_2018}. 

Motivated by these results, we construct a LBW ansatz for the massive Schwinger model by discretizing the continuum BW weight and applying it to the local terms of the lattice Hamiltonian. For a half-chain bipartition $A = \{0,\dots,\ell-1\}$, the BW theorem suggests a linear spatial weight proportional to the distance from the entangling cut. {For a lattice spacing $a$, this leads to the general structure}
{\begin{equation}
    \bar{H}_A^{\rm LBW} = a \sum_{n \in A} w(n)\, h_n,
\end{equation}}
where $h_n$ denotes the local Hamiltonian densities and $w(n)$ is a position-dependent weight increasing linearly with the distance from the cut.

For the Schwinger Hamiltonian in the spin formulation, the local contributions consist of mass terms, hopping terms, and electric-field terms. The corresponding LBW ansatz takes the explicit form
\begin{align}
    \bar{H}_A^{\rm LBW} &= \sum_{n \in A}^{\ell-1} w_{\text{bound}}(n) \, J_n(\theta)\, h_{n,n+1}^{\rm hop}\nonumber\\
    &+ \sum_{n \in A}^{\ell} w_{\text{site}}(n) \, J'_n(\theta)\, h_n^{\rm mass} \nonumber \\
    &+ \sum_{n \in A}^{\ell-1} w_{\text{bound}}(n) \, h_n^{E},
\end{align}
where {$w_{\text{site}}=a(\ell-n-1/2)$ and $w_{\text{bound}} =a( \ell-n-1) $} encode the linear distance from the entangling cut, and $J_n(\theta)$, $J'_n(\theta)$ inherit the $\theta$-dependent coefficients of the microscopic Hamiltonian \eqref{finalHam}. This construction mirrors the continuum BW expression, but with discretized weights adapted to the half-chain geometry.

To test the validity of this LBW construction, we compare it to the exact modular Hamiltonian obtained from the reduced density matrix. From exact diagonalization of the ground state $|\Psi_0\rangle$, we compute
\begin{equation}
    \rho_A = \mathrm{Tr}_B |\Psi_0\rangle\langle \Psi_0|,
\end{equation}
and diagonalize $\rho_A$ to obtain the exact entanglement eigenvalues and eigenvectors $\{|\psi_\alpha^{A}\rangle\}$.

Independently, we diagonalize $\bar{H}_A^{\rm LBW}$ to obtain its eigenvectors $\{|\psi_\alpha^{\rm EH}\rangle\}$. Agreement of the entanglement spectra provides a necessary consistency check. A stronger, operator-level test is obtained by computing the overlaps
\begin{equation}
    M_{\alpha,\alpha'} = \left| \langle \psi_\alpha^{\rm EH} | \psi_{\alpha'}^{A} \rangle \right|.
\end{equation}
A near diagonal overlap matrix indicates that the LBW ansatz approximately reproduces the eigenbasis of the exact modular Hamiltonian for the modes shown.

\section{Results}
We study the dependence of the massive Schwinger-model vacuum on the topological angle $\theta$ and the mass-to-coupling ratio $m/g$ using the ground-state energy density, chiral condensate, electric field, and bipartite entanglement observables. We first examine their $\theta$ dependence at three representative masses, then vary $m/g$ at fixed $\theta$ to characterize the near-critical region. Two-point correlation functions provide complementary information about the spatial extent of vacuum correlations. Finally, we test the lattice Bisognano--Wichmann ansatz by comparing its eigenvectors with those of the reduced density matrix.

Except for the LBW analysis in Sec.~\ref{Sec:LBW}, all the results are obtained using matrix product states (MPS), implemented with the \textit{ITensor} library~\cite{Fishman_2022,Corbett:2025flm} in the \textit{Julia} programming language~\cite{bezanson2015juliafreshapproachnumerical}. The $\theta$ scans use $N=1500$ staggered sites at $m/g=0.05$, $0.33$, and $0.42$, while the mass scan in Sec.~\ref{Sec:massdep} uses $N=1400$. All simulations are performed with open boundary conditions and $g=0.1$ (see~\cite{Grieninger:2026bdq} for a discussion of minimal lattice requirements in the Schwinger model).

A technical aspect of our simulations concerns the staggered lattice formulation. The $(-1)^n$ structure of the staggered mass term breaks invariance with respect to translation by a lattice spacing, and with OBC the discretization admits two distinct and equally valid even/odd sublattice conventions, although these conventions are equivalent in the continuum limit ($a\rightarrow0$).
To suppress these lattice artifacts, we compute all observables reported here for both staggered orientations and report their average. This symmetrization cancels the leading even/odd contributions and improves convergence toward the continuum $\theta$-vacuum physics.

The need for larger lattices at increasing $m/g$ originates from the fact that a $\theta$-term is equivalent to placing an external electric charge on the first and last lattice site, i.e. a background electric field whose strength grows with $\theta$. If the volume is too small, these charges may form a string state that looks like a phase transition/critical point but is purely induced by finite size effects. 
Moreover, using larger system sizes allows us to isolate bulk behavior away from the boundaries and ensures that the extracted observables are not distorted by edge effects inherent to open boundary conditions.

The LBW comparison in Sec.~\ref{Sec:LBW} is performed using exact diagonalization on a chain of $N=20$ sites. For a half-chain bipartition, we explicitly construct the reduced density matrix and compare its eigenvectors with those of the  LBW ansatz. This tests how well the ansatz reproduces the entanglement eigenbasis for the modes considered.

\begin{figure}
    \centering
    \includegraphics[width=1\linewidth]{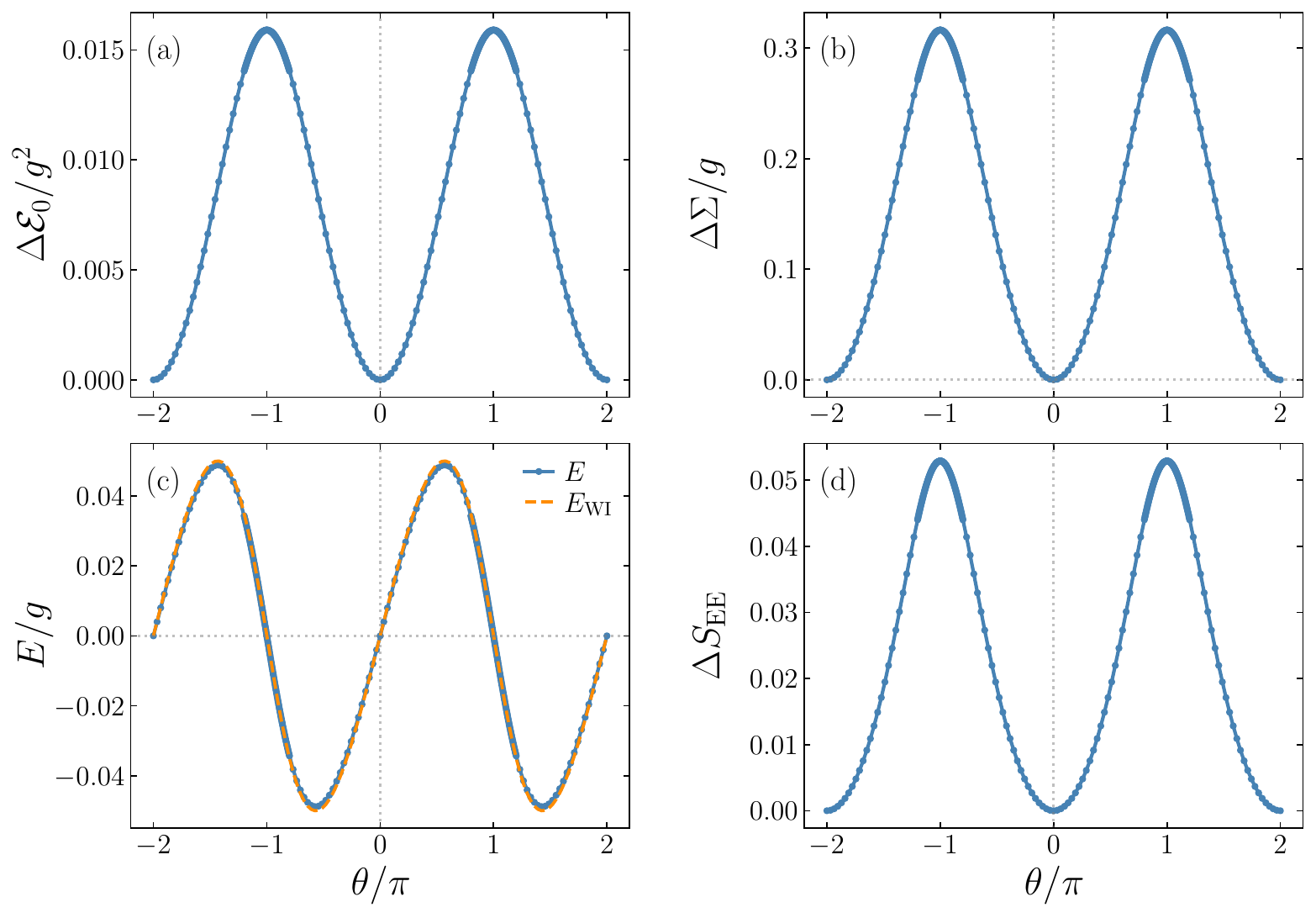}
    \caption{ 
    Ground-state observables of the massive Schwinger model as functions of the topological angle $\theta$ in the small-mass regime $m/g=0.05$ ($N=1500$). 
    {Panel (a) shows the ground-state energy-density difference 
    $\Delta \mathcal{E}_0(\theta)/g^2$, 
    (b) the chiral-condensate difference $\Delta\Sigma(\theta)/g=[\Sigma(\theta)-\Sigma(0)]/g$,
    (c) the spatially averaged electric field, $E/g$, together with the Ward-identity reconstruction, $E_{\rm WI}/g$, 
    and (d) the half-chain entanglement-entropy difference
    $\Delta S_{\rm EE}(\theta)=S_{\rm EE}(\theta)-S_{\rm EE}(0)$.
    The two electric-field estimators agree closely.
    All observables vary smoothly and exhibit $2\pi$-periodic
    behavior. The entropy difference develops broad maxima at
    $\theta=\pm\pi$.}
    }
    \label{GSenergy005}
\end{figure}

\subsection{The $\theta$ dependence of the vacuum structure}
\subsubsection{Ground-state energy}

\begin{figure}
    \centering
    \includegraphics[width=1\linewidth]{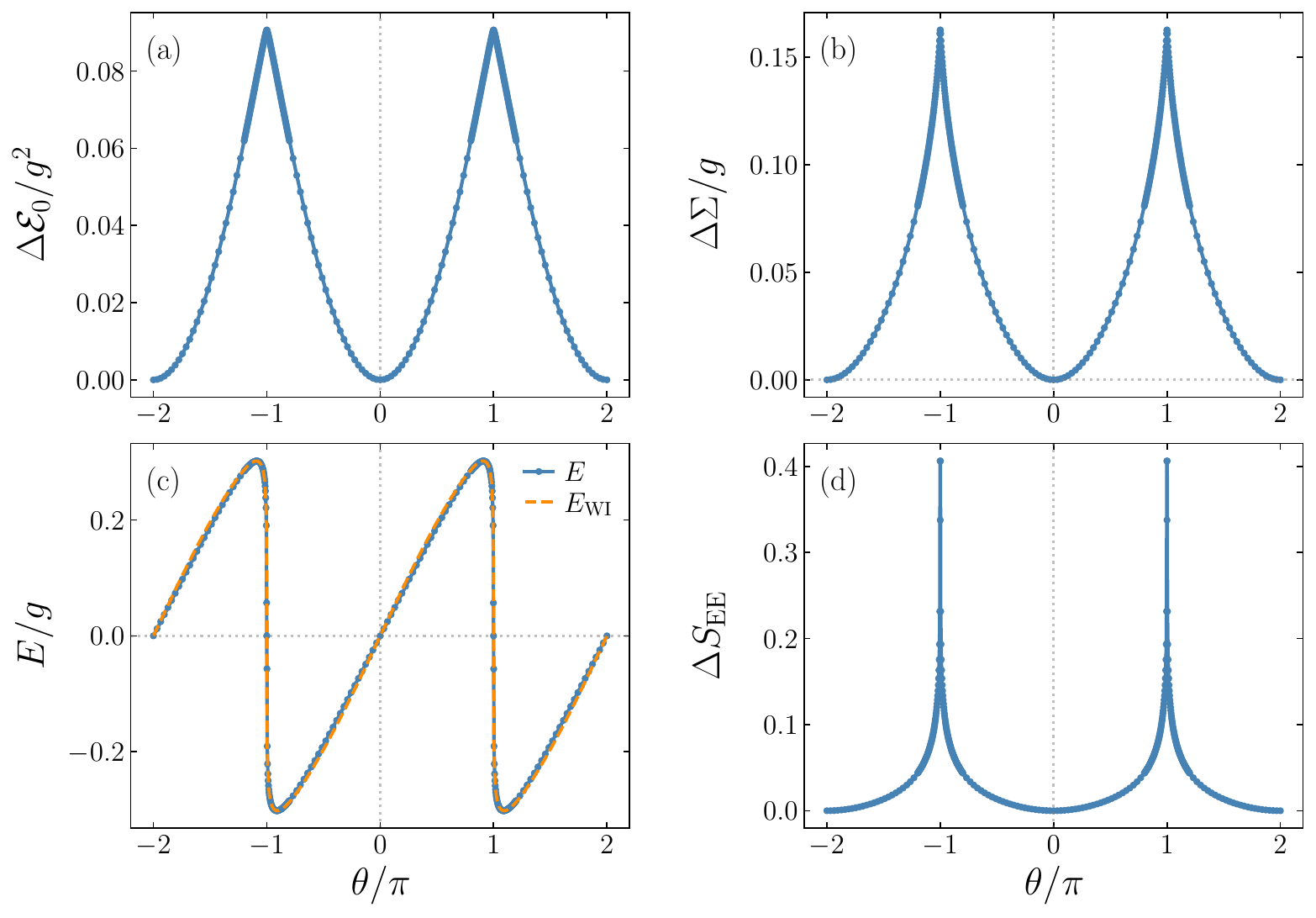}
    \caption{ Ground-state observables of the massive Schwinger model as functions of $\theta$ for $m/g=0.33$ ($N=1500$), close to the critical region. {Panel (a) shows the ground-state energy-density difference $\Delta\mathcal E_0(\theta)/g^2$, (b) the chiral-condensate difference $\Delta\Sigma(\theta)/g$, (c) the spatially averaged electric field obtained directly from the gauge operator, $E/g$, together with the Ward-identity reconstruction, $E_{\rm WI}/g$, and (d) the half-chain entanglement-entropy difference $\Delta S_{\rm EE}$. The two electric-field estimators are consistent with each other. Compared with the small-mass regime, the energy-density slope varies more sharply near $\theta=\pm\pi$, accompanied by a rapid electric-field reversal and a sharply localized enhancement of the entropy difference.}} 
    \label{GSenergy033}
\end{figure}

\begin{figure}
    \includegraphics[width=0.97\linewidth]{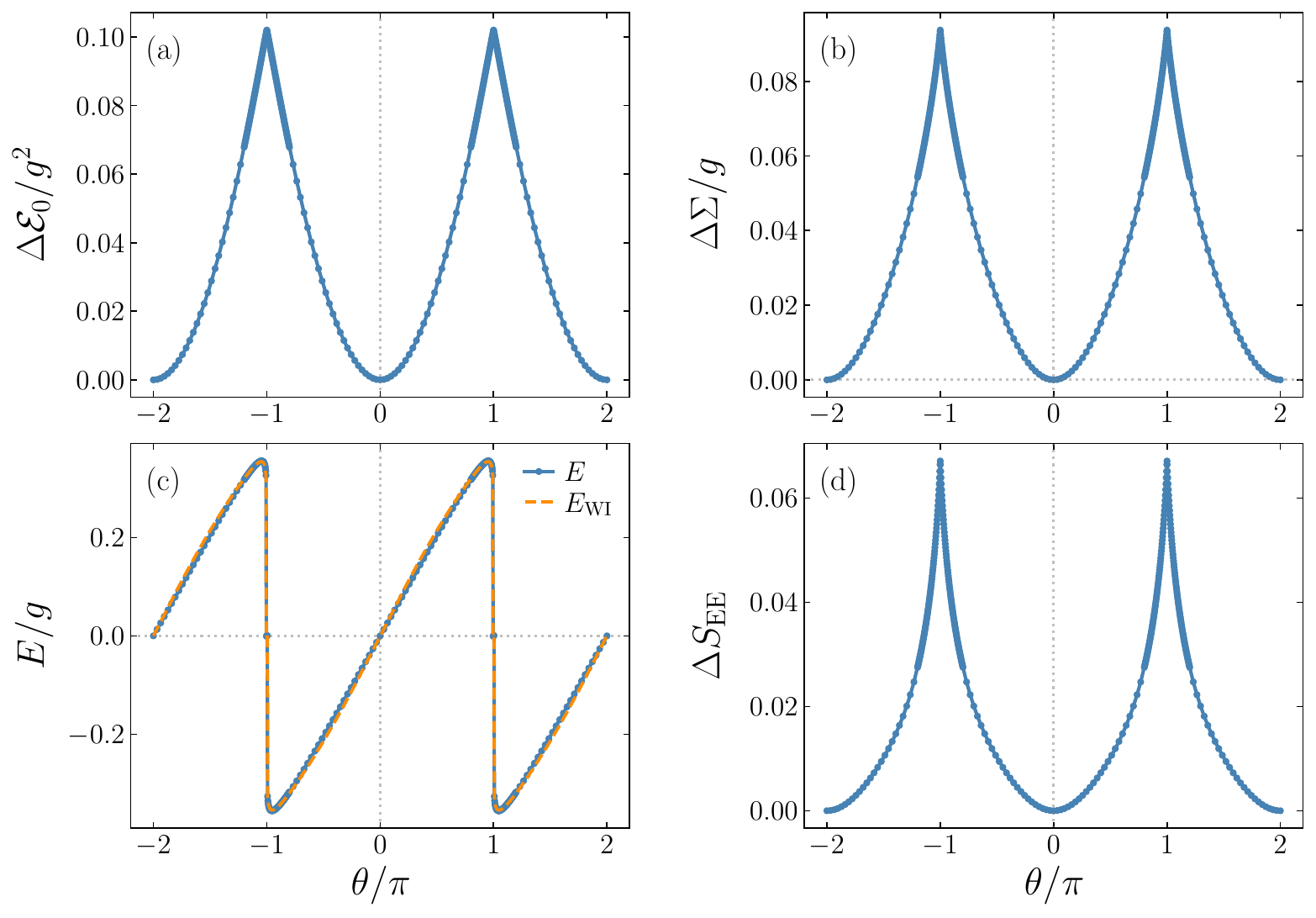}
    \caption{
    Ground-state observables of the massive Schwinger model as functions of $\theta$ for $m/g=0.42$ ($N=1500$).
    {Panel (a) shows the ground-state energy-density difference $\Delta\mathcal E_0(\theta)/g^2$,
    (b) the chiral-condensate difference $\Delta\Sigma(\theta)/g$,
    (c) the spatially averaged electric field obtained directly from the gauge operator, $E/g$, together with the Ward-identity reconstruction, $E_{\rm WI}/g$,
    and (d) the half-chain entanglement-entropy difference $\Delta S_{\rm EE}$.
    The two electric-field estimators remain consistent.
    The energy density retains a sharp maximum near
    $\theta=\pm\pi$, while the electric-field reversal becomes
    more abrupt. The entropy difference exhibits a narrow
    peak whose height is smaller than in the near-critical
    case $m/g=0.33$.}}
    \label{GSenergy04}
\end{figure}

The (a) panels of Figs. \ref{GSenergy005}, \ref{GSenergy033} and \ref{GSenergy04} show the ground-state energy density difference $ \Delta \mathcal E_0(\theta)/g^2=[\mathcal E_0(\theta)- \mathcal E_0(0)]/g^2$ obtained from the lattice Hamiltonian. 
The lattice ground-state energy is extensive, so comparison with the continuum energy density requires division by the physical length $Na$~\cite{LenaPRD2020}.

In addition, the chirally rotated discretization introduces the $\theta$-dependent contact term derived in Sec.~\ref{contact_terms}. Denoting the unsubtracted lattice ground-state eigenvalue by $E_0^{\rm raw}(\theta)$, the quantity reported here is
\begin{equation}
    \frac{\Delta\mathcal E_0(\theta)}{g^2}
    =
    \frac{E_0^{\rm raw}(\theta)-E_0^{\rm raw}(0)}{g^2Na}
    -
    \frac{c_a^{(0)}(\theta)}{g^2},
\end{equation}
where $c_a^{(0)}(0)=0$.
Although the energy requires no observable rotation, this contact-term subtraction is necessary for comparison with the continuum prediction; it is not removed by subtracting the value at $\theta=0$.


For comparison, we use the continuum mass-perturbation expansion of Refs.~\cite{ADAM19971,ADAM1998117}, expressed in the conventions $\Sigma=\partial_m\mathcal E_0$ and $E=(2\pi/g)\partial_\theta\mathcal E_0$. In the small-mass regime, $m/g\ll1$, the energy density through second order in $m/g$ is, up to a mass- and $\theta$-independent constant,
\begin{align}
\frac{\mathcal E_0(m,\theta)}{g^2}
&= -\frac{m\Sigma_0}{g^2}\cos\theta
- \pi \left(\frac{m\Sigma_0}{2g^2}\right)^2 \nonumber \\
&\times \left( \mu_{0E+}^2 \cos 2\theta
+ \mu_{0E-}^2 \right),
\end{align}
where $\Sigma_0=ge^\gamma/(2\pi^{3/2})$ is the positive chiral-condensate scale, $\gamma$ is the Euler--Mascheroni constant, and $\mu_{0E+}^2=-8.9139$ and $\mu_{0E-}^2=9.7384$ are the coefficients of the continuum expansion. 

Subtracting the value at $\theta=0$ yields the UV-finite expression
\begin{align}
\frac{\Delta \mathcal E_0(\theta)}{g^2}
&= \frac{m\Sigma_0}{g^2}\bigl(1-\cos\theta\bigr)
+\pi \left(\frac{m\Sigma_0}{2g^2}\right)^2 \nonumber \\
&\times \mu_{0E+}^2 \bigl(1-\cos 2\theta\bigr).
\label{GSenergy_chiral}
\end{align}
The leading contribution therefore predicts a smooth cosine
dependence, with maxima at odd multiples of $\pi$.

For $m/g=0.05$, the numerical energy difference follows the smooth, approximately cosine dependence predicted by Eq.~\eqref{GSenergy_chiral}.
Appendix~\ref{App:A} compares the small-mass results for the energy density, condensate, and electric field with their continuum predictions at several lattice spacings.

At $m/g=0.33$ and $0.42$, the energy maximum near $\theta=\pi$ becomes sharper, and the angular dependence departs from the small-mass prediction. The cusp-like profile indicates the breakdown of the perturbative description and is consistent with the expected first-order transition at $\theta=\pi$ in the thermodynamic limit.

\subsubsection{Chiral condensate}
The (b) panels of Figs.~\ref{GSenergy005}, \ref{GSenergy033}, and~\ref{GSenergy04} show the spatially averaged condensate difference $\Delta\Sigma(\theta)=\Sigma(\theta)-\Sigma(0)$.
We evaluate the rotated operator in Eq.~\eqref{Eq:chiralcondensate}, divide the lattice sum by $N$, and apply the contact-term subtraction described in Sec.~\ref{contact_terms}. The subtraction of the $\theta=0$ reference then gives the UV-finite difference used for comparison with the continuum prediction.

In the continuum, the condensate is the mass derivative
of the ground-state energy density,
\begin{equation}
\Sigma(m,\theta) = \frac{\partial \mathcal E_0(m,\theta)}{\partial m},
\label{eq:condpert}
\end{equation}
and therefore measures the response of the vacuum energy
to changes in the fermion mass. Differentiating the
small-mass expansion gives~\cite{ADAM19971,ADAM1998117}
\begin{align}
\frac{\Sigma(m,\theta)}{g}
&= -\frac{\Sigma_0}{g}\cos\theta
- \frac{\pi m}{2g}\left(\frac{\Sigma_0}{g}\right)^2 
\left(\mu_{0E+}^2 \cos 2\theta + \mu_{0E-}^2 \right).
\end{align}
At leading order, the condensate difference is thus
$\Delta\Sigma(\theta)=\Sigma_0(1-\cos\theta)$.

For $m/g=0.05$, the numerical results closely follow this smooth angular dependence, with broad maxima near $\theta=\pm\pi$. At $m/g=0.33$ and $0.42$, the peaks become sharper, while their heights decrease relative to the small-mass case. The condensate difference therefore exhibits a stronger localization of its angular variation near $\theta=\pm\pi$, rather than a monotonic increase in its magnitude.

\subsubsection{Electric field}

The spatially averaged electric field probes the response of the vacuum energy density to the topological angle. In the continuum,
\begin{equation}
    E(m,\theta)
    =
    \frac{2\pi}{g}
    \frac{\partial\mathcal E_0(m,\theta)}{\partial\theta}.
    \label{eq:elfield}
\end{equation}

On the lattice, we evaluate the electric field using the two estimators introduced in Sec.~\ref{chirally_rot_obs}. The gauge-field estimator $E$ is obtained by spatially averaging $\langle E_n\rangle=g\langle L_n\rangle$, where $L_n$ is determined by Gauss's law. The second estimator, $E_{\rm WI}$, is reconstructed from the anomalous Ward identity using fermionic bilinears in the chirally rotated basis.

Their comparison provides a consistency check of the gauge-field and fermionic definitions. The contact-term subtraction of Sec.~\ref{contact_terms} is applied to $E_{\rm WI}$, whereas the direct gauge-field estimator requires no such subtraction. The resulting estimators agree closely in the (c) panels of Figs.~\ref{GSenergy005}--\ref{GSenergy04}.

In the small-mass regime, differentiating the continuum
energy density yields
\begin{align}
\frac{E(m,\theta)}{g}
&= 2\pi\,\frac{\partial}{\partial\theta}\!\left(\frac{\mathcal E_0(m,\theta)}{g^2}\right) \nonumber \\
&= 2\pi\,\frac{m\Sigma_0}{g^2}\sin\theta
+\pi^2\left(\frac{m\Sigma_0}{g^2}\right)^2 \mu_{0E+}^2 \sin(2\theta).
\end{align}
The leading angular dependence is sinusoidal, with higher harmonics suppressed by additional powers of $m/g$. For $m/g=0.05$ [Fig.~\ref{GSenergy005}(c)], both estimators follow this smooth behavior, consistent with the perturbative
continuum prediction.

For $m/g=0.33$ [Fig.~\ref{GSenergy033}(c)], the electric field departs from the nearly sinusoidal small-mass behavior and changes rapidly near $\theta=\pi$. Through Eq.~\eqref{eq:elfield}, this behavior corresponds to the sharp variation of the energy-density slope in the near-critical regime.

For $m/g=0.42$ [Fig.~\ref{GSenergy04}(c)], the reversal of the electric field becomes more abrupt, consistent with switching between opposite-electric-flux branches in the regime where a first-order transition is expected in the thermodynamic limit. The gauge-field and Ward-identity estimators remain consistent across this sharper variation.

\begin{figure}
    \centering
    \includegraphics[width=1\linewidth]{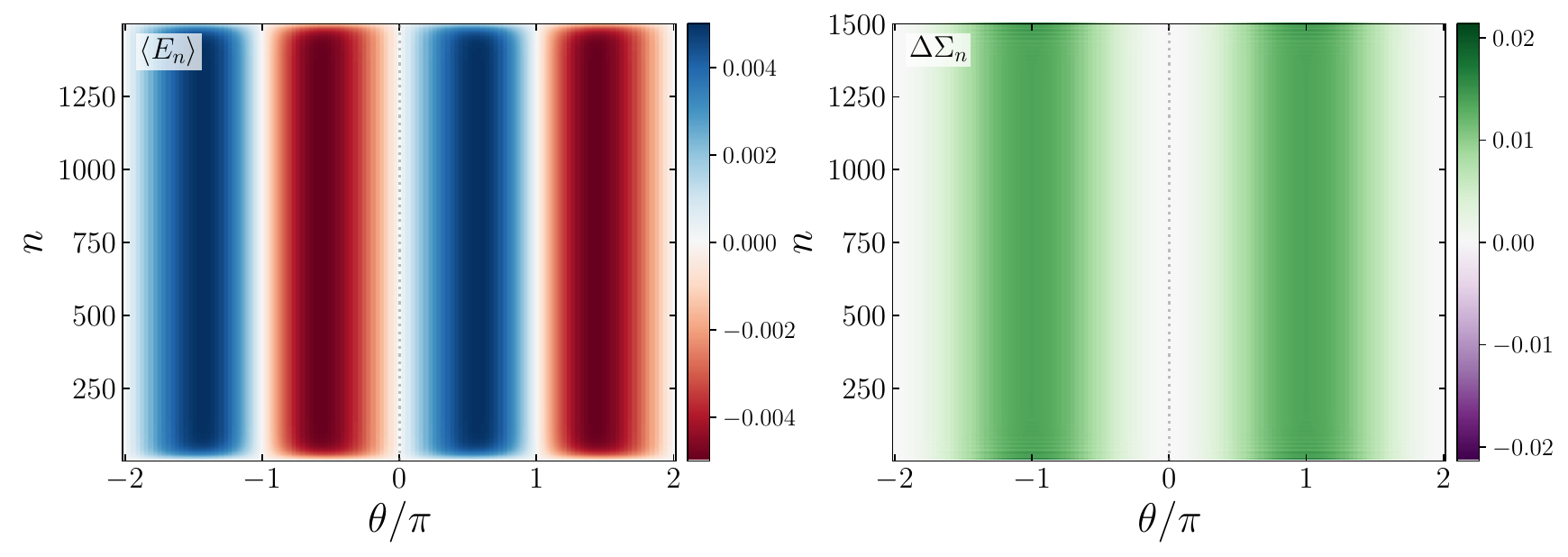}
    \includegraphics[width=1\linewidth]{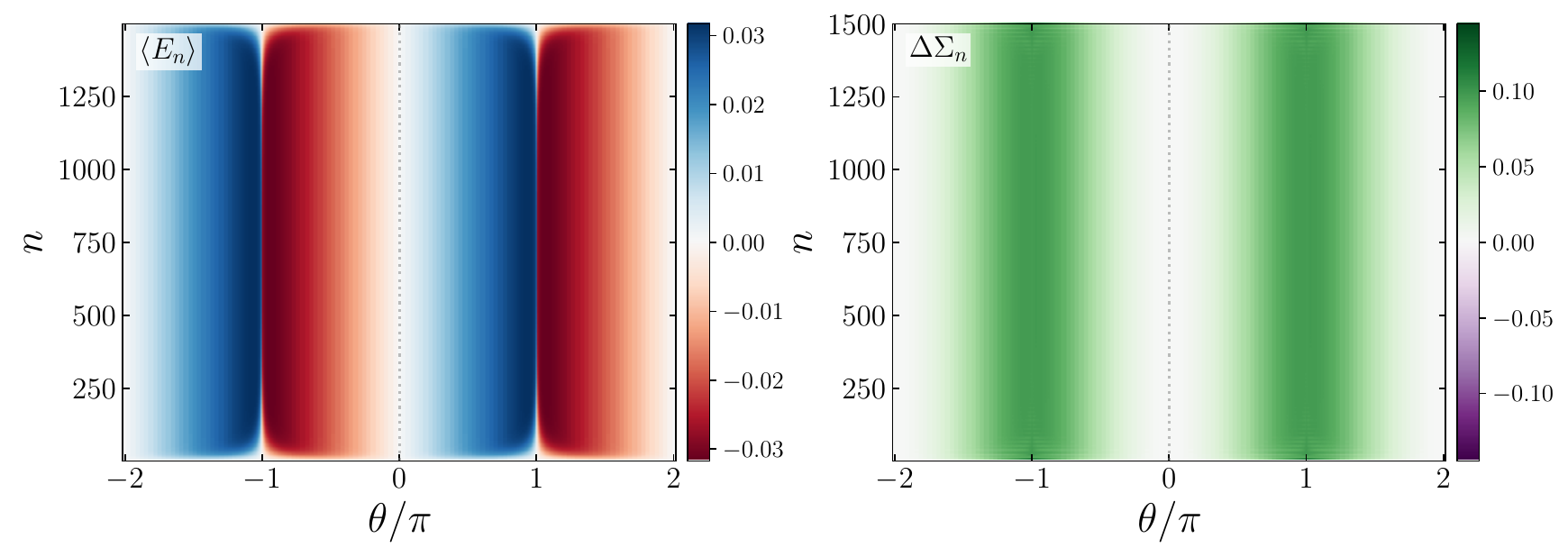}
    \includegraphics[width=1\linewidth]{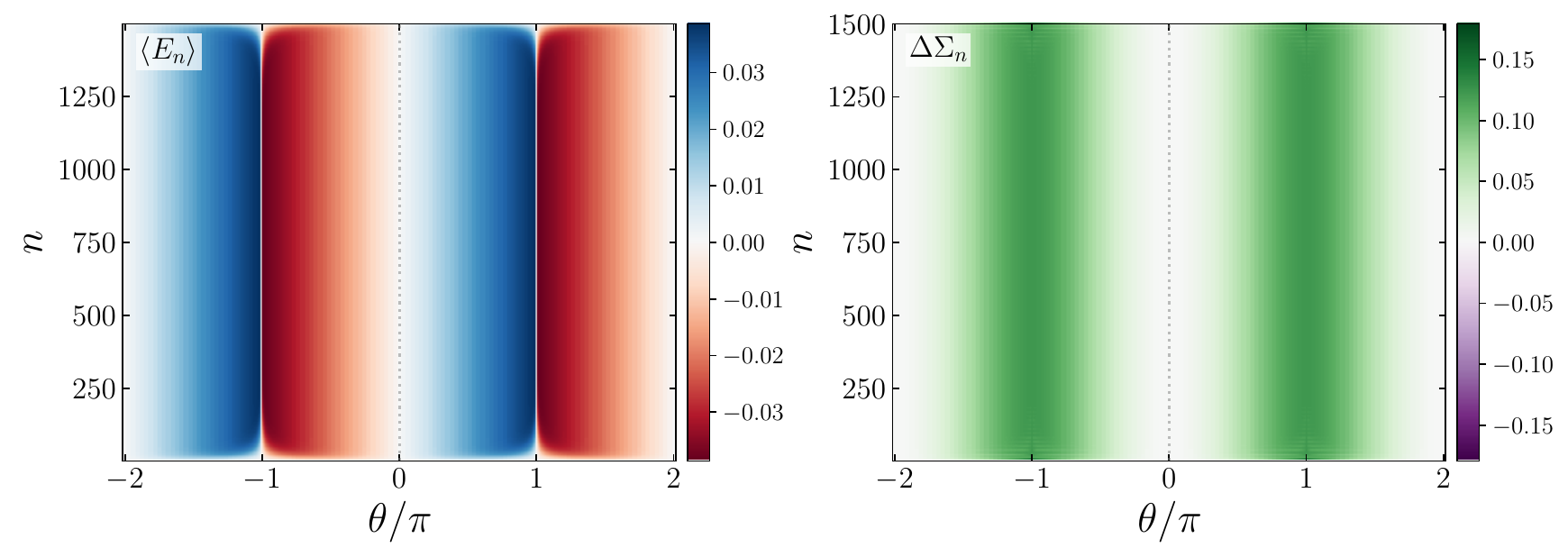}
    \caption{
    Spatial dependence of vacuum observables for fermion masses
    $m/g=0.05$ (top), $m/g=0.33$ (middle), and $m/g=0.42$ (bottom),
    for system size $N=1500$.
    Left: local electric field $\langle E_n\rangle$ as a function of
    $\theta$ and lattice position $n$.
    Right: local condensate difference
    $\Delta\Sigma_n(\theta)=\Sigma_n(\theta)-\Sigma_n(0)$.
    Both observables are approximately uniform throughout the bulk, with spatial variations confined mainly to the boundaries, demonstrating the effective recovery of bulk translational invariance in the large open chains.
    }
    \label{spatial_profiles}
\end{figure}

{The spatially resolved observables are shown in
Fig.~\ref{spatial_profiles} for the three representative masses $m/g=0.05$, $0.33$, and $0.42$. The left panels show the local electric field $\langle E_n\rangle$, while the right panels show the local condensate difference $\Delta\Sigma_n(\theta)=\Sigma_n(\theta)-\Sigma_n(0)$. At fixed $\theta$, the observables are approximately uniform in the interior, with the largest position dependence near the boundaries. These profiles support interpreting the measured angular dependence as representative of the bulk over the
central region of the chains studied.}

{The electric field is odd under $\theta\to-\theta$, with the orientation of the electric flux reversing as the sign of $\theta$ is changed, whereas the condensate difference is even under this transformation. These spatial profiles therefore confirm that the $\theta$-dependent features discussed above are properties of the bulk vacuum and not boundary-induced effects.}

\begin{figure}
    \centering
 \includegraphics[width=1\linewidth]{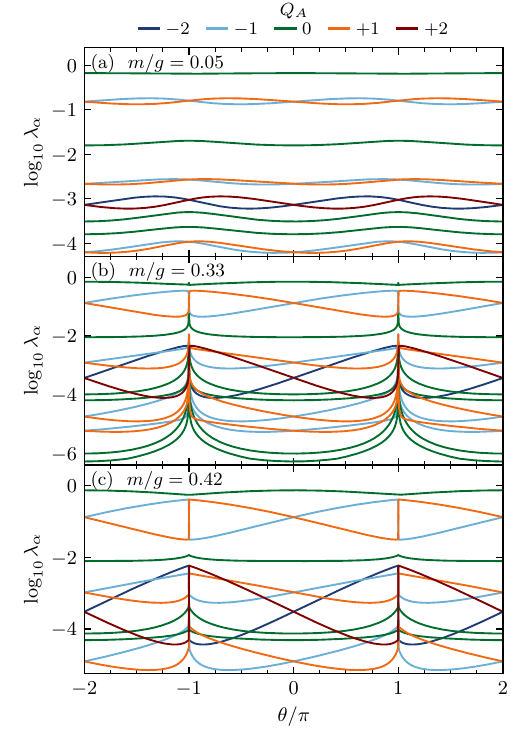}
    \caption{
    {Symmetry-resolved Schmidt weights of the half-chain reduced density matrix as functions of $\theta$ for (a) $m/g=0.05$, (b) $m/g=0.33$, and (c) $m/g=0.42$. The leading weights are displayed as $\log_{10}\lambda_\alpha$ over $\theta/\pi\in[-2,2]$, with colors indicating the subsystem-charge sectors $Q_A$. At $\theta=0,\pm\pi,\pm2\pi$, corresponding weights in opposite-charge sectors are degenerate for all three masses. The pronounced compression of distinct low-lying entanglement levels near $\theta=\pm\pi$ is most evident for $m/g=0.33$.}
    }
    \label{schmidt_all}
\end{figure}

\subsubsection{Entanglement entropy and entanglement spectrum}

We characterize half-chain entanglement using the von Neumann entropy and the symmetry-resolved entanglement spectrum.
The entropy difference $\Delta S_{\rm EE}(\theta)=S_{\rm EE}(\theta)-S_{\rm EE}(0)$ is shown in the (d) panels of Figs.~\ref{GSenergy005}, \ref{GSenergy033}, and~\ref{GSenergy04}, while the largest Schmidt weights are displayed in Fig.~\ref{schmidt_all}.

The entropy difference exhibits maxima near $\theta=\pm\pi$ for all three masses, with a shape and height that depend strongly on $m/g$. For $m/g=0.05$, the enhancement is broad and smooth. Near the critical region, at $m/g=0.33$, the peak becomes sharply localized and substantially larger.
For $m/g=0.42$, the peak remains narrow, but its height is considerably reduced. Thus, an entropy enhancement near $\theta=\pi$ occurs across the masses studied, while its magnitude is greatest in the near-critical case.

{The symmetry-resolved spectra in Fig.~\ref{schmidt_all} provide complementary information. At $\theta=0,\pm\pi,\pm2\pi$, corresponding Schmidt weights in opposite-charge sectors $Q_A=q$ and $Q_A=-q$, with $q\neq0$, are degenerate since the total electric field vanishes there. This pairing occurs for all three masses and therefore does not by itself identify the critical region.}

{For $m/g=0.05$, the spectrum varies smoothly with $\theta$, with relatively weak changes in the separations between distinct levels. At $m/g=0.33$, several low-lying entanglement levels become markedly compressed near $\theta=\pm\pi$ (see Fig.~\ref{schmidt_all}(b)), resulting in the peak of EE. At $m/g=0.42$, the entanglement spectrum shows the first order phase transition but the overall spectrum narrows less compared to $m/g=0.33$. Among the three masses studied, the near-critical case is therefore distinguished by the combination of a larger entropy enhancement and a more pronounced narrowing of the low-lying entanglement spectrum.}

\subsection{Mass dependence and the critical region}
\label{Sec:massdep}
We study the mass dependence using the chirally rotated Hamiltonian~\eqref{finalHam} at $\theta=0$ with $N=1400$. 
The scalar and pseudoscalar mass terms are proportional to $m\cos\theta$ and $m\sin\theta$, respectively. 
Consequently, a negative mass at $\theta=0$ gives the same mass terms as a positive mass of equal magnitude at $\theta=\pi$.
The negative-mass scan therefore probes the regime associated with the critical point of the conventional positive-mass theory at $\theta=\pi$.

\begin{figure*}
    \centering
    \includegraphics[width=1\linewidth]{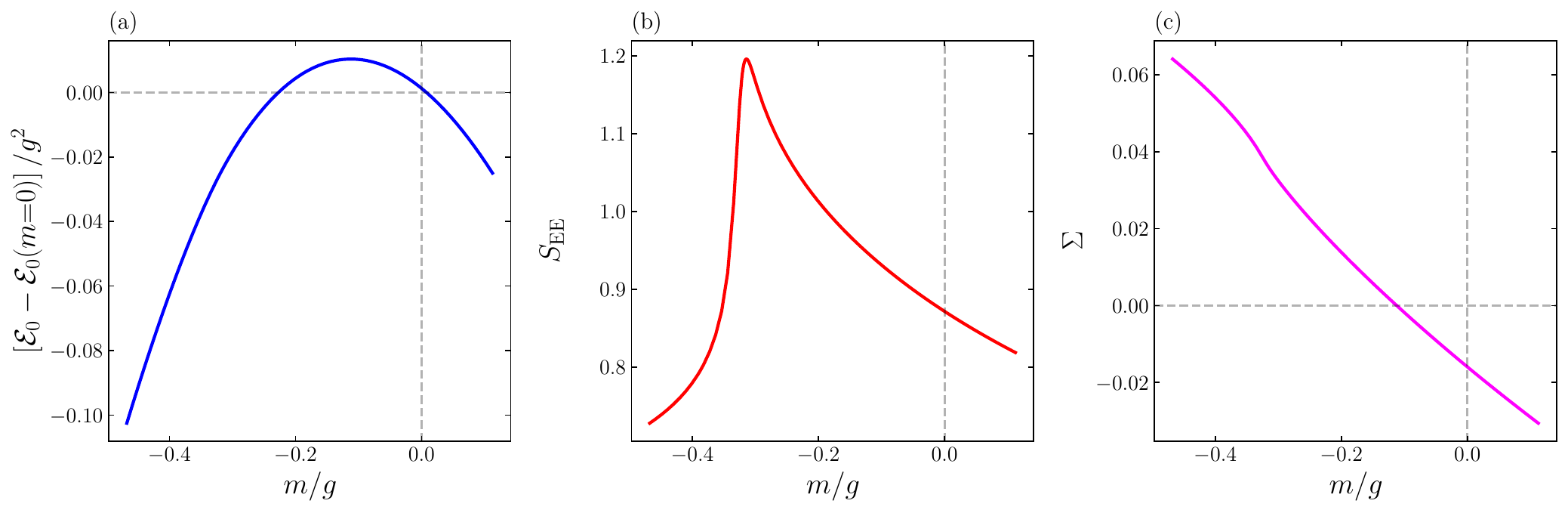}
    \caption{
    Mass dependence of vacuum observables for $N=1400$ in the chirally rotated formulation at $\theta=0$.
    Negative masses correspond to the positive-mass theory at $\theta=\pi$.
    Panels show (a) the ground-state energy-density difference $[\mathcal E_0(m)-\mathcal E_0(m{=}0)]/g^2$, (b) the half-chain entanglement entropy $S_{\rm EE}$, and (c) the spatially averaged chiral condensate $\Sigma$. The entropy exhibits a pronounced maximum at $m/g\simeq-0.314$.
    }

    \label{obs_mg}
\end{figure*}

Figure~\ref{obs_mg} shows the the ground-state energy-density difference, the half-chain entanglement entropy, and spatially averaged chiral condensate as functions of $m/g$. 
The {energy-density difference $[\mathcal E_0(m)-\mathcal E_0(m{=}0)]/g^2$} varies smoothly and exhibits a broad maximum.
The condensate also varies smoothly over the mass range shown. 
{In contrast, the entanglement entropy develops a clear maximum at $m/g\simeq-0.314$. 
This maximum lies near the critical region associated with the conventional theory at $\theta=\pi$, whose critical mass ratio is approximately $m/g\simeq0.33$~\cite{Byrnes_2002,Byrnes_2002-2,CruzPRD025}.}
Thus, among the observables shown in Fig.~\ref{obs_mg}, the entanglement entropy provides the clearest signature of the critical region. 

\begin{figure}
    \centering
    \includegraphics[width=1\linewidth]{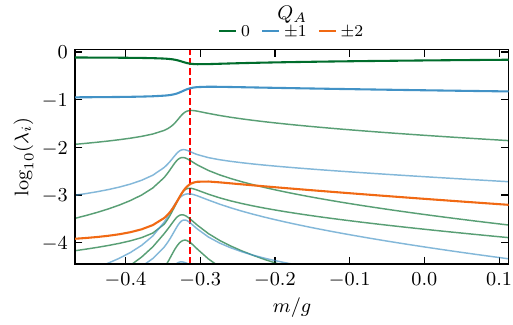}
    \caption{
    {Mass dependence of the symmetry-resolved half-chain entanglement spectrum at $\theta=0$ for $N=1400$. The leading Schmidt weights are displayed as $\log_{10}\lambda_i$, with colors distinguishing the subsystem-charge sectors $Q_A=0,\pm1,\pm2$. Corresponding weights in opposite-charge sectors coincide throughout the scan.
    The vertical red dashed line marks the half-chain entropy maximum at $m/g\simeq-0.314$, near which several distinct low-lying entanglement levels become compressed.}
    }
    \label{schmidt_mg}
\end{figure}

{Figure~\ref{schmidt_mg} shows the corresponding symmetry-resolved entanglement spectrum. At $\theta=0$, corresponding Schmidt weights in opposite-charge sectors $Q_A=q$ and $Q_A=-q$, with $q\neq0$, remain degenerate throughout the mass scan.  This pairing therefore does not locate the critical region. The mass-dependent restructuring is instead visible in the compression of several distinct low-lying entanglement levels near the entropy maximum.}

After averaging over the two staggered sublattice conventions, the spatially averaged electric field vanishes throughout the mass range. 
This follows from the symmetry of the Hamiltonian at $\theta=0$, where the two lattice orientations correspond to opposite electric-field configurations and their average removes any net flux. 
The absence of a net electric field further illustrates that the restructuring near the critical mass is more clearly resolved by entanglement observables than by the local expectation values considered here.

Together, the entropy enhancement and spectral compression identify a mass range near the expected critical region. We next examine two-point correlation functions in Sec.~\ref{subsec:corr} to determine how these entanglement features relate to the growth of spatial correlations.


\subsubsection{Topological susceptibility}
We further characterize the vacuum response through the topological susceptibility, defined as the curvature of the ground-state energy density with respect to $\theta$. We report the dimensionless ratio
\begin{equation}
    \frac{\chi_{\rm top}(m)}{g^2}
    =
    \left.
    \frac{\partial^2}{\partial\theta^2}
    \frac{\mathcal E_0(m,\theta)}{g^2}
    \right|_{\theta=0}.
\end{equation}
The susceptibility $\chi_{\rm top}$ itself has dimensions of mass squared. Its relation to the integrated connected correlator of the topological charge density provides the corresponding Euclidean interpretation~\cite{Meggiolaro_1998,Mao_2009}.

We evaluate this curvature numerically while varying $m/g$, using the energy density after the contact-term subtraction described in Sec.~\ref{contact_terms}. Through Eq.~\eqref{eq:elfield}, the susceptibility also determines the response of the electric field to a small change in $\theta$:
\begin{equation}
    \chi_{\rm top}(m)=
    \frac{g}{2\pi} \left. \frac{\partial E(m,\theta)}{\partial\theta} \right|_{\theta=0}.
\end{equation}

For $|m|/g\ll1$, differentiating Eq.~\eqref{GSenergy_chiral} gives
\begin{equation}
    \frac{\chi_{\rm top}(m)}{g^2} =
    \frac{m\Sigma_0}{g^2}
    +\pi\left(\frac{m\Sigma_0}{g^2}\right)^2
    \mu_{0E+}^2
    +\mathcal O\!\left((m/g)^3\right).
\end{equation}
The leading contribution is therefore linear in $m$, with a sign determined by the mass parameter.

\begin{figure}
    \centering
    \includegraphics[width=0.9\linewidth]{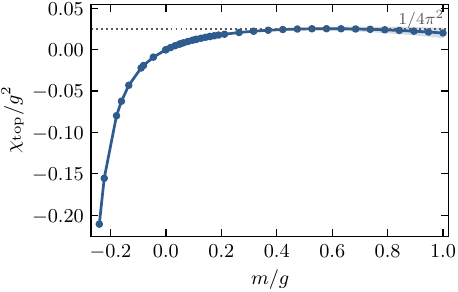}
    \caption{
    {Dimensionless topological susceptibility
    $\chi_{\rm top}/g^2$ as a function of $m/g$, obtained from the curvature of the ground-state energy density with respect to $\theta$ at $\theta=0$, after the contact-term subtraction. For small positive $m/g$, the susceptibility increases from zero, consistently with the leading behavior $\chi_{\rm top}\propto m$. At larger positive masses, the ratio approaches the pure-gauge value $1/(4\pi^2)$, indicated by the horizontal dotted line. For negative $m/g$, the susceptibility is negative and its magnitude grows as the mass becomes more negative over the range shown.}
    }
    \label{top_mg}
\end{figure}

{The numerical results are shown in Fig.~\ref{top_mg}. For small positive $m/g$, the susceptibility increases from zero, consistently with the leading perturbative behavior. As $m/g$ increases, the dimensionless ratio $\chi_{\rm top}/g^2$ approaches the pure-gauge value $1/(4\pi^2)$, indicated by the horizontal dotted line.
In the heavy-fermion limit, screening is suppressed and the vacuum energy approaches
\begin{equation}
    \frac{\mathcal E_0(\theta)-\mathcal E_0(0)}{g^2}
    =
    \frac{\theta^2}{8\pi^2},
    \qquad |\theta|<\pi,
\end{equation}
on the branch containing $\theta=0$. Consequently,
\begin{equation}
    \lim_{m/g\to+\infty}
    \frac{\chi_{\rm top}}{g^2}
    =
    \frac{1}{4\pi^2}.
\end{equation}
The numerical trend is consistent with this crossover from the light-fermion regime toward weak coupling.}

{For negative $m/g$, corresponding to the conventional positive-mass theory at $\theta=\pi$, the susceptibility is negative and its magnitude grows rapidly as the mass becomes more negative over the range shown. This behavior reflects an increasing curvature magnitude of the vacuum energy with respect to $\theta$ as the critical region is approached.}

{The small-mass behavior follows from the chiral expansion of the vacuum energy density in Eq.~\eqref{GSenergy_chiral}. Taking its second derivative with respect to $\theta$ and evaluating at $\theta=0$ gives the perturbative prediction for $\chi_{\rm top}$ as a function of $m/g$. This provides a reference for $|m|/g\ll1$, including the leading sign change across $m=0$. The growth in the magnitude of the susceptibility at increasingly negative masses indicates an enhanced vacuum response to $\theta$.}

\begin{figure*}
    \centering
    \includegraphics[width=0.96\linewidth]{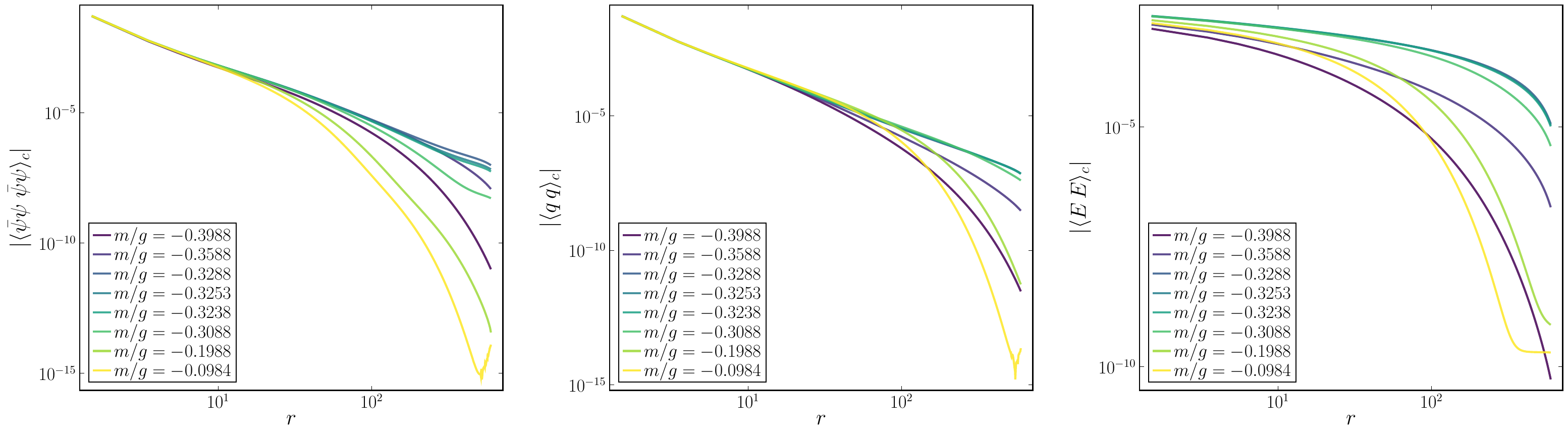}
    \caption{
    Spatial decay of connected two-point correlators at $\theta=0$ for $N=1400$ and several mass ratios $m/g$. From left to right, the panels show $|C_{\bar\psi\psi}(r)|$, $|C_q(r)|$, and $|C_E(r)|$, corresponding to the scalar density, charge density, and electric field. The separation $r$ is measured from the reference site $n_0=N/2$ in staggered lattice sites. Both axes are logarithmic.
    }
    \label{corr_func}
\end{figure*}
\begin{figure}
    \centering
    \includegraphics[width=1\linewidth]{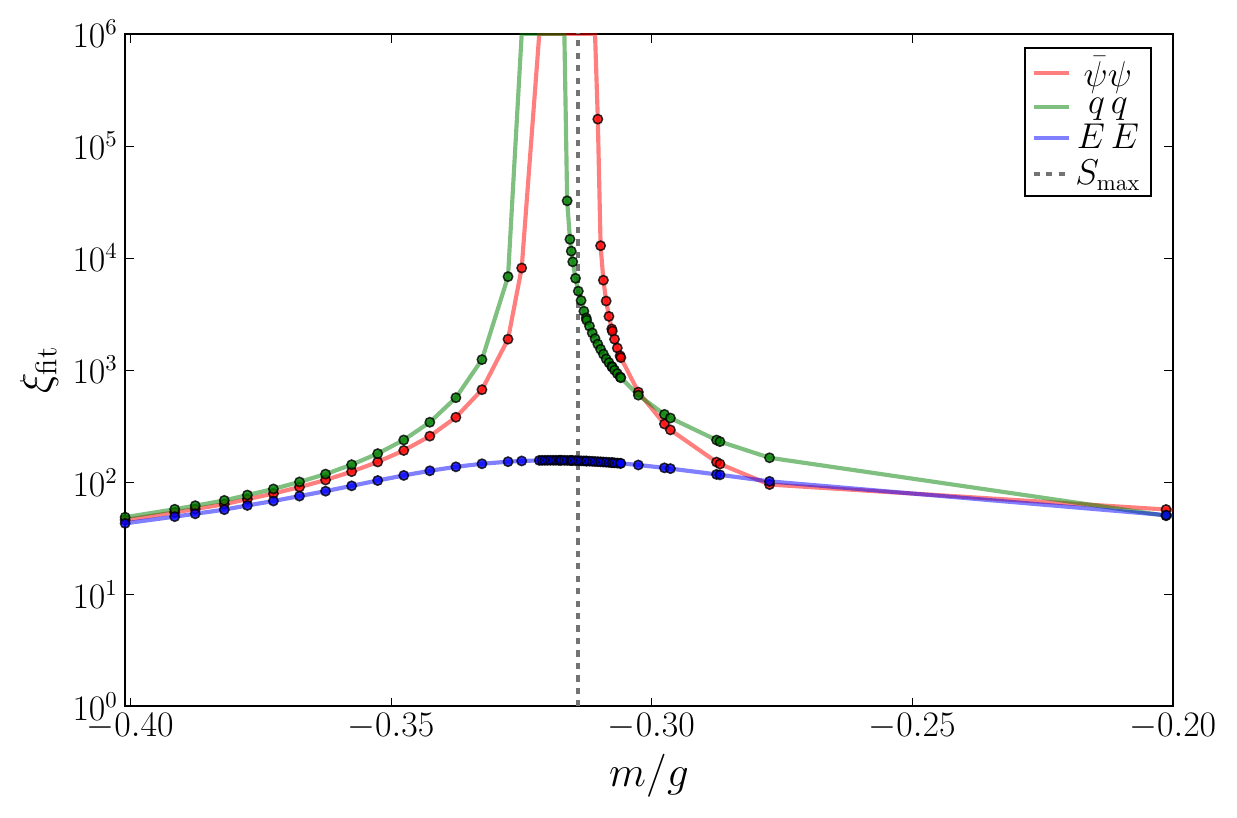}%
    \caption{
    Fitted correlation lengths $\xi_O$ as functions of $m/g$ at $\theta=0$ for $N=1400$, extracted from the connected scalar-density, charge-density, and electric-field correlators. The black dash-dotted line marks the half-chain entropy maximum at $m/g\simeq-0.314$. 
    {Near $m/g\simeq-0.33$, the fitted lengths in the scalar-density and charge channels grow beyond the accessible separations, where their numerical values are poorly constrained. The electric-field channel exhibits a more moderate enhancement. The large fitted values indicate an unresolved exponential decay scale on the finite lattice.}
    }
    \label{corr_length}
\end{figure}

\subsection{Correlation function and correlation length}
\label{subsec:corr}

To complement the entanglement observables, we examine spatial correlations on a lattice with $N=1400$ sites at $\theta=0$. {We define the connected correlators
\begin{equation}
    C_O(r)
    =
    \langle O_{n_0}O_{n_0+r}\rangle
    -
    \langle O_{n_0}\rangle\langle O_{n_0+r}\rangle,
\end{equation}}
where $n_0=\{N/2-1,N/2\}$ is the physical site at midpoint and $O$ denotes the scalar density $\bar\psi\psi$, the charge density $q$, or the electric field $E$. Here, $r$ is the separation from the reference site, measured in staggered lattice sites.

Figure~\ref{corr_func} shows the magnitudes of these correlators on logarithmic axes. Away from the near-critical mass range, their decay is relatively rapid. Near $m/g\simeq-0.33$, the slower decay indicates an enhancement of correlations over large separations.

{We extract a correlation length for each channel by fitting the large-distance decay to
\begin{equation}
    |C_O(r)|
    \simeq
    A_O r^{-\eta_O}\exp(-r/\xi_O),
\end{equation}
where $A_O$, $\eta_O$, and $\xi_O$ are fit parameters.} To reduce boundary effects, the fit range excludes points within 100 staggered sites of the boundary, corresponding to 50 two-site cells. The resulting fitted correlation lengths are shown in Fig.~\ref{corr_length}.

Near the critical region, the fitted lengths in the charge and scalar-density channels become much larger than the accessible separations. In this regime, the exponential decay scale is poorly constrained, and the correlators are compatible with algebraic decay over the fitted range. This occurs for the charge correlator over
$m/g\in[-0.3328,-0.3221]$ and for the scalar-density correlator over $m/g\in[-0.3361,-0.3279]$. These intervals characterize the finite-size resolution of the extraction.

The fitted electric-field correlation length shows a more moderate enhancement and remains finite within our numerical resolution. Away from the near-critical region on either side, the fitted lengths decrease and the exponential decay becomes more readily resolved.

{The strongest growth of the fitted correlation lengths occurs near $m/g\simeq-0.33$. Together with the peak in the EE, the correlation functions support the identification of a near-critical region with enhanced long-distance correlations.}

\subsection{Relation between the energy gap and the entanglement spectrum}
\label{Sec:LBW}

We investigate the connection between the low-energy structure of the massive Schwinger model and the ground-state half-chain entanglement spectrum using the lattice Bisognano--Wichmann framework.

\begin{figure}
    \centering
    \includegraphics[width=\linewidth]{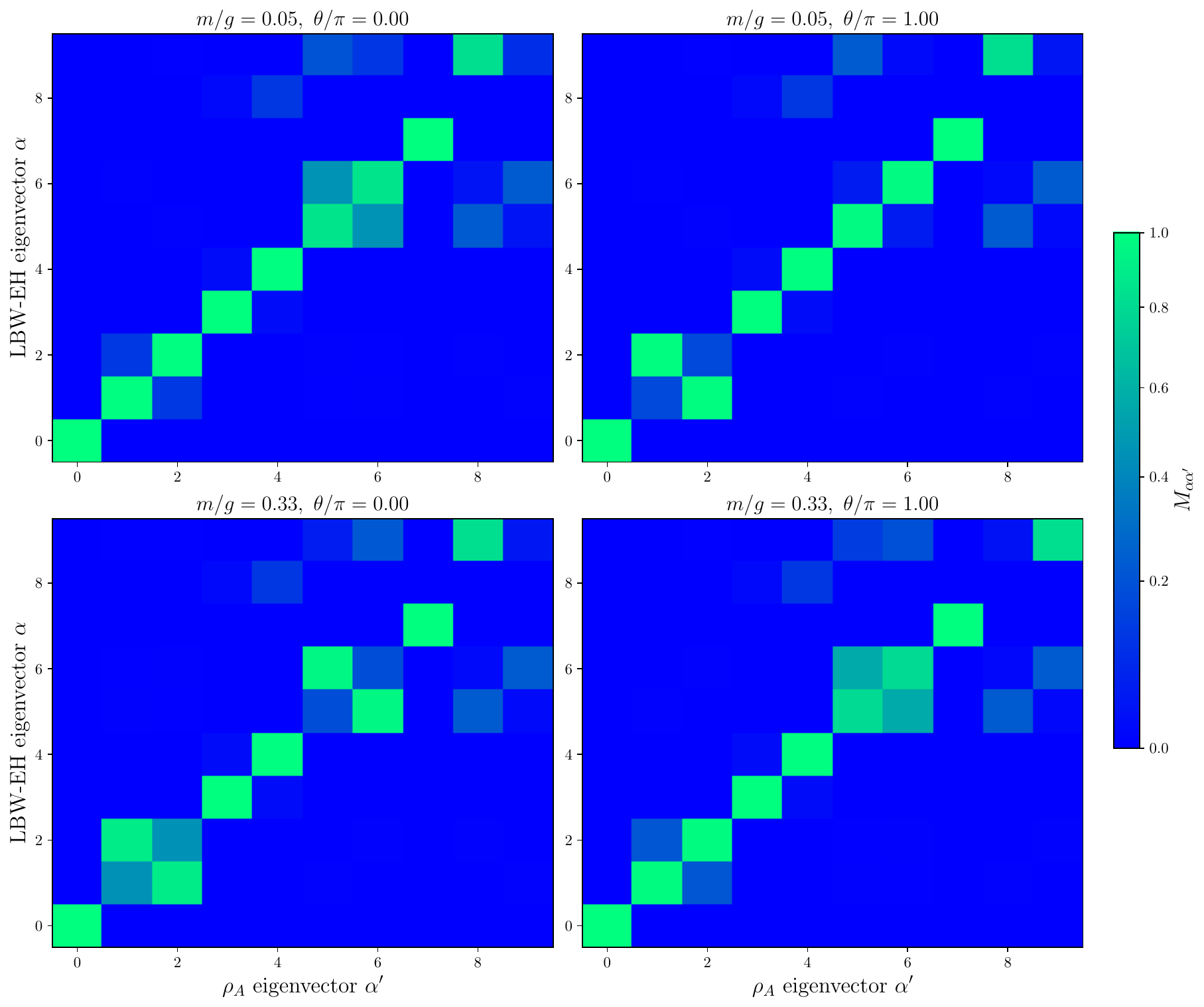}
    \caption{
    Absolute overlap matrix
    $M_{\alpha,\alpha'} = \left|\langle\psi^{\rm EH}_\alpha |\psi^{A}_{\alpha'}\rangle\right|$
    between eigenvectors of the lattice Bisognano--Wichmann entanglement Hamiltonian and the exact reduced density matrix for a half-chain bipartition ($N=20$, $\ell=10$). Results are shown for two representative mass ratios $m/g=0.05$ and $0.33$, and for $\theta=0$ and $\pi$. The predominantly near-diagonal structure for the leading modes indicates that the LBW ansatz approximately reproduces the corresponding eigenvectors of the exact modular Hamiltonian. This eigenvector agreement supports an approximate BW description on the finite lattice.
    }
    \label{fig:LBW_overlap}
\end{figure}

The connection is naturally understood within the Bisognano--Wichmann framework discussed in Sec.~II. For a half-space bipartition of a relativistic vacuum, the modular Hamiltonian has an exact local expression constructed from the physical Hamiltonian density with a spatially varying weight. Our lattice BW analysis tests whether this construction approximately reproduces the eigenvectors associated with the lowest entanglement energies, corresponding to the largest eigenvalues of the reduced density matrix. This agreement is illustrated in Fig.~\ref{fig:LBW_overlap}, where the overlap matrix between the LBW eigenvectors and the exact entanglement eigenvectors exhibits a predominantly near-diagonal structure for the leading modes across the values of $m/g$ and $\theta$ considered.

From this perspective, variations in the physical excitation gap may be reflected in the structure of the entanglement spectrum. The LBW ansatz is constructed from spatially weighted terms of the microscopic Hamiltonian, providing a connection between the operators governing physical excitations and the entanglement structure.
The eigenvector comparison supports this connection, although it does not by itself establish agreement of the entanglement eigenvalues or of the full modular Hamiltonian.

Near a continuous critical point, a decreasing physical gap is associated with a growing correlation length and more spatially extended low-energy fluctuations.
{Within the BW framework, this restructuring of the infrared sector is consistent with the compression of low-lying entanglement levels observed near criticality. The present comparison supports a qualitative connection between the entanglement spectrum and the physical low-energy structure, without establishing a quantitative relation between a particular entanglement-level spacing and the excitation gap.}

\section{Discussion}
In this work, we investigate how the $\theta$-vacuum structure of the massive Schwinger model is captured by local observables and bipartite entanglement. Our results connect the angular dependence of the electric field and vacuum energy with spatial correlations and the charge-resolved entanglement spectrum. In particular, they distinguish the entropy enhancement observed near $\theta=\pi$ across the masses studied from the more pronounced spectral compression found near criticality.

The $\theta$ term was implemented through a chirally rotated lattice formulation, which makes its $2\pi$ periodicity explicit at the Hamiltonian level.
An important result of this work is the identification and subtraction of the $\theta$-dependent vacuum-energy artifact introduced by this discretization.
We derive its exact free-lattice form and apply the corresponding subtractions to the energy density and its derivatives. Together with the lattice improvement discussed in Sec.~II, this treatment enables consistent comparisons with continuum predictions.

At small $m/g$, the chiral condensate, ground-state energy density, and electric field agree closely with the analytic chiral perturbation results.
The agreement between the gauge-field and Ward-identity estimators provides an additional consistency check of the rotated formulation and the subtraction procedure. At larger $m/g$, the angular dependence becomes sharper near $\theta=\pi$, with a more abrupt reversal of the electric field, consistent with competition between opposite-electric-flux vacuum branches.

{Entanglement observables provide complementary information about this vacuum structure. The entropy difference is enhanced near $\theta=\pi$ for all three masses studied, whereas the near-critical case $m/g\simeq0.33$ exhibits a substantially larger enhancement and a pronounced compression of the low-lying entanglement spectrum. Resolving the spectrum by subsystem charge distinguishes this collective restructuring from the degeneracies between opposite-charge sectors at $\theta=0,\pi$, which also occur away from criticality. This distinction clarifies which spectral features provide information beyond the presence of charge-sector pairing.}

{The mass scan provides complementary signatures of the critical region where we show that several distinct entanglement levels become compressed near the half-chain entropy maximum at $m/g\simeq-0.314$. The fitted correlation lengths in the scalar-density and charge channels grow sharply near $m/g\simeq-0.33$, while the electric-field channel shows a more moderate enhancement. Where the fitted lengths exceed the accessible separations, their values cannot be reliably resolved on the finite chain. The diverging correlation length and the peak in the EE occur near the known quantum critical point. Finite-size and continuum scaling of the entropy and charge-resolved spectra would establish how these features converge and whether their separation persists.}

The lattice Bisognano--Wichmann construction provides a concrete link between the microscopic Hamiltonian and the entanglement structure of the vacuum.
For the finite chains studied by exact diagonalization, the LBW ansatz approximately reproduces the eigenvectors associated with the leading Schmidt weights. This agreement shows that a spatially weighted Hamiltonian built from known microscopic operators captures the dominant entanglement eigenbasis, providing a physically motivated description of this structure without reconstructing the full modular Hamiltonian.
It also motivates exploring the LBW ansatz as a route to accessing entanglement properties through Hamiltonian engineering in quantum simulations. The present comparison establishes eigenvector agreement; testing the entanglement eigenvalues and extending the analysis to larger systems would determine the accuracy and range of applicability of this description.

More generally, the results illustrate how the entropy and spectrum encode different aspects of vacuum restructuring. The entropy summarizes the redistribution of Schmidt weights, while the charge-resolved spectrum identifies the sectors involved and distinguishes persistent degeneracies from changes in the separation of distinct levels. Their combined analysis provides a more informative characterization of the near-critical vacuum than either observable alone.

A related perspective comes from topological phases, where the entanglement spectrum can reveal information about boundary excitations that is not captured by the entropy alone. In fractional quantum Hall systems, the low-lying entanglement-level structure and counting have been related to those of physical edge modes \cite{PhysRevLett.101.010504,PhysRevB.84.205136}. These results illustrate how the entanglement spectrum can encode physically meaningful excitation structure even when the bulk is gapped. In the Schwinger model, the charge-resolved spectral restructuring near $\theta=\pi$ similarly motivates examining how competing electric-flux configurations are represented in the entanglement degrees of freedom. Comparing boundary conditions and the flux content of the Schmidt states would clarify whether a more direct connection to boundary excitations can be established.

Quantum-wire platforms offer a complementary route to exploring connections between entanglement and measurable charge fluctuations. For noninteracting fermions coupled through a quantum point contact, Klich and Levitov related the generated entanglement entropy to the full counting statistics of charge transfer, involving current fluctuations and their higher cumulants \cite{Klich_2009}. This provides an operational way to access entanglement without reconstructing the full reduced density matrix.
Motivated by this connection, it would be interesting to investigate whether tunable junctions or phase-biased topological wires exhibit related changes in charge fluctuations and entanglement when low-energy configurations compete. Establishing such a correspondence could provide an experimentally accessible analogue of aspects of the vacuum
restructuring studied here.
\newline

\begin{acknowledgments}

This work was supported by the U.S. Department of Energy, Office of Science, Office of Nuclear Physics, Grants No. DE-SC0020970 (Inqubator for Quantum Simulation (IQuS), S.G.). This work was also supported by the U.S. Department of Energy, Office of Science, Office of Nuclear Physics, Grant No. DE-FG02-97ER-41014 (UW Nuclear Theory, S.G.), DE-FG02-88ER40388 (SBU Nuclear Theory, D.K., E.M.) and by the U.S. Department of Energy, Office of Science, National Quantum Information Science Research Centers, Co-design Center for Quantum Advantage (C2QA) under Contract No.DE-SC0012704 (D.K., E.M.). S.G. was supported in part by a Feodor Lynen Research fellowship of the Alexander von Humboldt foundation. E.M. was supported in part by the Center for Distributed Quantum Processing at Stony Brook University. 
This work was also supported, in part, by the Department of Physics and the College of Arts and Sciences at the University of Washington.
This work was enabled, in part, by the use of advanced computational, storage and networking infrastructure provided by the Hyak supercomputer system at the University of Washington.
\end{acknowledgments}

\bibliographystyle{apsrev4-1}
\bibliography{apssamp}

\appendix

\newpage
\section{Continuum limit}
\label{App:A}

We approach the continuum limit using a fixed-volume procedure in which $Nag$ is held constant while increasing $N$ and decreasing $a$, thereby isolating discretization effects at fixed physical scale.
We compute the ground-state expectation values of the chiral condensate, the ground-state energy density, and the electric field, and compare their behavior across different lattice spacings. This provides a direct test of convergence toward the continuum limit.

\subsection{Numerical results}

\begin{figure}[ht]
    \centering \includegraphics[width=\linewidth]{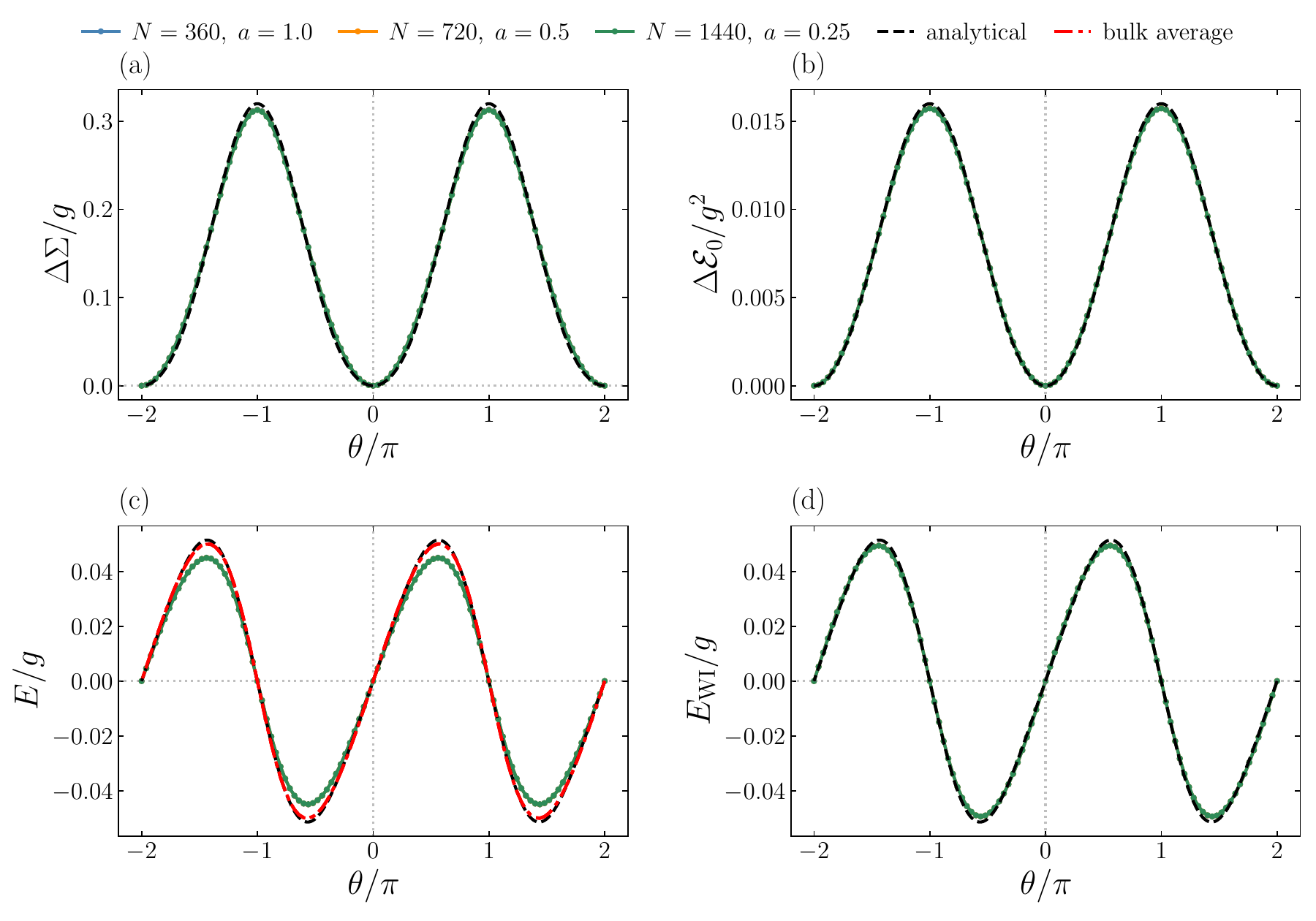}
    \caption{
    Continuum-limit study at fixed physical volume $Nag=36$ for $m/g=0.05$. 
    We compare three lattice discretizations with $(N,a)=(360,1.0)$, $(720,0.5)$, and $(1440,0.25)$ (colored lines), together with the analytical continuum prediction (black dashed line). 
    (a) Difference in the chiral condensate $\Delta\Sigma/g$. 
    (b) Ground-state energy density difference $\Delta \mathcal E_0/g^2$. 
    (c) Spatially averaged electric field $E/g$ obtained from the gauge operator. 
    (d) Electric field $E_{\rm WI}/g$ reconstructed from the anomalous Ward identity in Eq.~\eqref{Eq:avgEfield}. 
    The curves show weak dependence on lattice spacing at fixed physical volume. The full-chain electric-field average retains a boundary contribution in panel (c), which is reduced by averaging over the bulk.}    
    \label{fig:contlimit}
\end{figure}

In particular, at $\theta=0$ the Hamiltonian is CP symmetric, while $\bar{\psi}i\gamma_5\psi$ is CP odd. Its expectation value therefore vanishes, which implies $E=0$. Away from $\theta=0$, the chiral rotation generates a nonzero CP-odd combination of fermion bilinears, leading to a nonvanishing average electric field.

In Fig.~\ref{fig:contlimit}, the agreement across different lattice spacings confirms that the fixed-volume scaling procedure reliably reproduces the continuum behavior. In particular, the collapse of the curves demonstrates that both the condensate and the electric field have reached their continuum form within numerical precision. 

Comparing the two definitions of the electric field, we find that the Ward-identity reconstruction $E_{\rm WI}$ exhibits slightly improved agreement with the perturbative continuum prediction, while the direct lattice definition based on the gauge operator remains fully consistent within the numerical precision of our results. This provides a nontrivial consistency check of the implementation of the $\theta$ dependence.
We emphasize that this convergence is achieved only after incorporating the improved Hamiltonian in Eq.~\eqref{finalHam}, which ensures the correct realization of the underlying symmetries and significantly reduces lattice artifacts.

\section{Correlation length}
In the main text the correlation length is extracted from the exponential decay of the spatial two-point correlators at large separations. 
As an additional consistency check, we also compute the second-moment correlation length $\xi_{\rm 2nd}$, which provides a fit-independent estimator obtained directly from the correlator.

For a correlator $C(r)$ as a function of lattice separation $r$ from the middle site, the second-moment correlation length is defined as
\begin{equation}
\xi_{\rm 2nd}^2 =
\frac{\sum_r r^2 C(r)}{2 \sum_r C(r)} ,
\end{equation}
which characterizes the spatial spread of correlations without relying on a specific large-distance fit window.

\begin{figure}[h]
    \centering
    \includegraphics[width=0.49\textwidth]{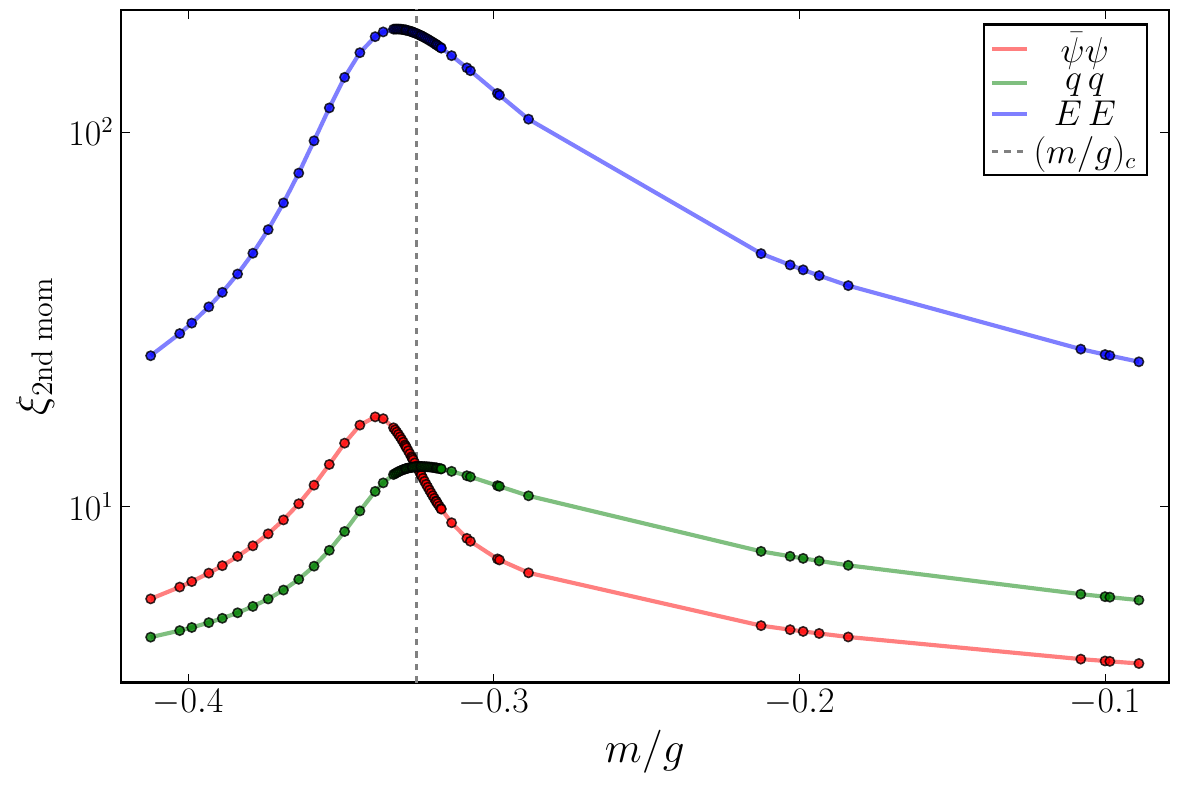}
    \caption{
    Second-moment correlation length $\xi_{\rm 2nd}$ as a function of the mass ratio $m/g$ for a lattice of size $N=1400$. 
    The curves correspond to the fermion bilinear correlator $\langle \bar{\psi}\psi\,\bar{\psi}\psi \rangle_c$, the charge--charge correlator $\langle q q \rangle_c$, and the electric-field correlator $\langle E E \rangle_c$. 
    The vertical dashed line indicates the peak value $(m/g)_c \simeq -0.314$ of the entanglement entropy $S_{\mathrm{max}}$. 
    A strong enhancement of $\xi_{\rm 2nd}$ is observed near this region, consistent with the correlation-length growth extracted from exponential fits in Fig.~12.
    }
    \label{2ndmomXi}
\end{figure}

The resulting values of $\xi_{\rm 2nd}$ are shown in Fig.~\ref{2ndmomXi} for several connected correlators. 
Although this estimator differs from the exponential fit used in the main analysis, it displays the same qualitative behavior: a pronounced enhancement of the correlation length near the region of $m/g \simeq -0.314$. 
This agreement confirms that the growth of correlations observed in the main text is robust and not an artifact of the fitting procedure.

\end{document}